# Molecular-Sized Outward-Swinging Gate: Experiment and Theoretical Analysis of a Locally Nonchaotic Barrier

Yu Qiao,[1,2,*] Zhaoru Shang,[1] Rui Kou[2]

[1] *Program of Materials Science and Engineering, University of California – San Diego, La Jolla, CA 92093, U.S.A.*

[2] *Department of Structural Engineering, University of California – San Diego, La Jolla, CA 92093-0085, U.S.A.*

*\* Corresponding author. Email: yqiao@ucsd.edu*

**ABSTRACT:** We investigate the concept of molecular-sized outward-swinging gate, which allows for entropy decrease in an isolated system. The theoretical analysis, the Monte Carlo simulation, and the direct solution of governing equations all suggest that under the condition of local nonchaoticity, the probability of particle crossing is asymmetric. It is demonstrated by an experiment on a nanoporous membrane one-sidedly surface-grafted with bendable organic chains. Remarkably, through the membrane, gas spontaneously and repeatedly flows from the low-pressure side to the high-pressure side. While this phenomenon seems counterintuitive, it is compatible with the principle of maximum entropy. The locally nonchaotic gate interrupts the probability distribution of the local microstates, and imposes additional constraints on the global microstates, so that entropy reaches a nonequilibrium maximum. Such a mechanism is fundamentally different from Maxwell's demon and Feynman's ratchet, and is consistent with microscopic reversibility. It implies that useful work may be produced in a cycle from a single thermal reservoir. A generalized form of the second law of thermodynamics is proposed.

# 1. Introduction

In an ergodic and chaotic system, when the particle distribution of an ideal gas is uniform, entropy ($S$) is maximized.[1] However, a nonchaotic or nonergodic system may not reach thermodynamic equilibrium.[2-7] A recent computer simulation discovered that a locally nonchaotic energy barrier could break the symmetry of the cross-influence of thermally correlated thermodynamic driving forces.[8]

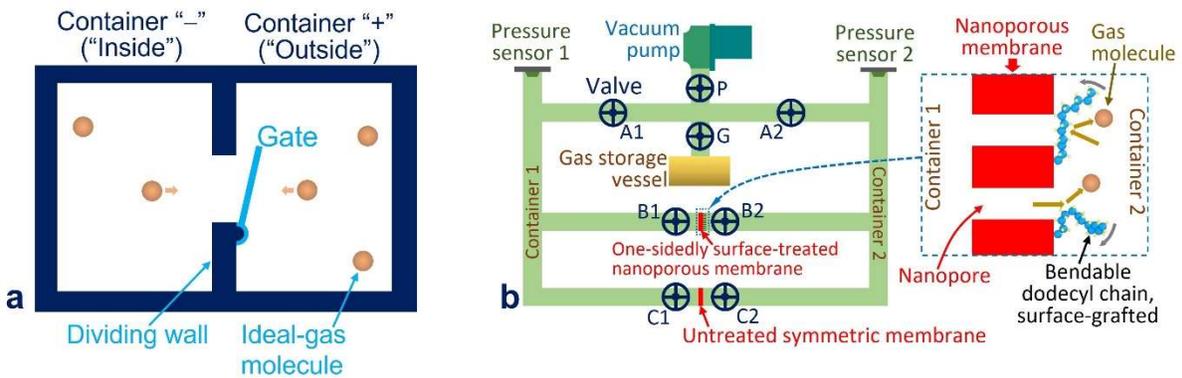

**Figure 1**. **(a)** Schematic of an isolated system consisting of two large containers filled with an ideal gas. The containers are connected through a small nanopore. A molecular-sized outward-swinging gate is at the side of container "+" ("outside"). **(b)** Schematic of the experimental setup. In some tests, the untreated membrane between valves C1 and C2 is replaced by a non-permeable solid film. The inset at the right-hand side shows a magnified view of the dodecyl chains surface-grafted at the nanopore openings.

In the current research, we investigate a locally nonchaotic entropy barrier, as depicted in Figure 1(a). In an isolated system, two large containers, "+" and "−", are filled with an ideal gas. The containers are connected through a small nanopore in the dividing wall. The system is initially at thermodynamic equilibrium; that is, the gas distribution is uniform in the two containers. There is a molecular-sized outward-swinging gate at the "+" side ("outside"). The nanopore size and the gate size are much smaller than the mean free path of the gas molecules, so that the gas molecules interact with the gate individually. The gate cannot cross the dividing wall; i.e., its swinging motion is limited in container "+". In the ideal-case scenario, the gate is rigid and lightweight. There may be a non-dissipative attraction force ($F_G$) between the gate and the "door stopper" on the dividing wall, which tends to trap the gate in a closed configuration. As will be discussed in Section 2, under the condition of local nonchaoticity, the overall crossing ratio of the gate, $\kappa = \delta_\pm/\delta_\mp$, does not equal to 1, where $\delta_\pm$ and $\delta_\mp$ are the probabilities for the gas molecules to cross the nanopore from container "−" to "+" and from container "+" to "−", respectively. As a result, the steady-state gas molecular density in container "+" is not the same as in container "−".

The second law of thermodynamics states that entropy of an isolated system cannot decrease.[1] Yet, when the isolated system in Figure 1(a) evolves from the uniform initial state to the nonuniform steady state, the distribution of the gas molecules spontaneously becomes nonequilibrium, and entropy is reduced. Below, in Sections 2-4, we consider the fundamental mechanism of the molecular-sized outward-swinging gate. In Sections 5 and 6, we experimentally demonstrate the concept by using a nanoporous membrane one-sidedly surface-grafted with bendable organic chains.

## 2. Microscopic reversibility: Why $\kappa \neq 1$

In this section, we show how the overall crossing ratio of the gate can be asymmetric ($\kappa \neq 1$): While every particle trajectory is time-reversible, the probabilities of the microstates associated with the forward process and the reverse process are different. The critical factors include the local nonchaoticity and the thermal motion of the gate.

### 2.1 Probability distribution of microstates

Figure 2(a) depicts the particle crossing event at an outward-swinging gate. For the forward process from left to right, the microstate of incident particle in container "−" is denoted by $\Psi_a$, and the microstate of outgoing particle in container "+" is denoted by $\Psi_b$. When the particle is at $\Psi_a$ and $\Psi_b$, the particle velocities are respectively denoted by $\vec{v}_a$ and $\vec{v}_b$, and the microstates of the gate are respectively denoted by $\Phi_a$ and $\Phi_b$. For the reverse process from right to left, $\overline{\bullet}$ indicates the reverse microstates. Compared with the forward microstate, the velocity direction of the reverse microstate is inverted, with everything else being identical.

It is worth noting that $\Psi_a$ and $\Psi_b$ are only for the crossing cases; if the particle is blocked by the gate, its microstates are not registered.

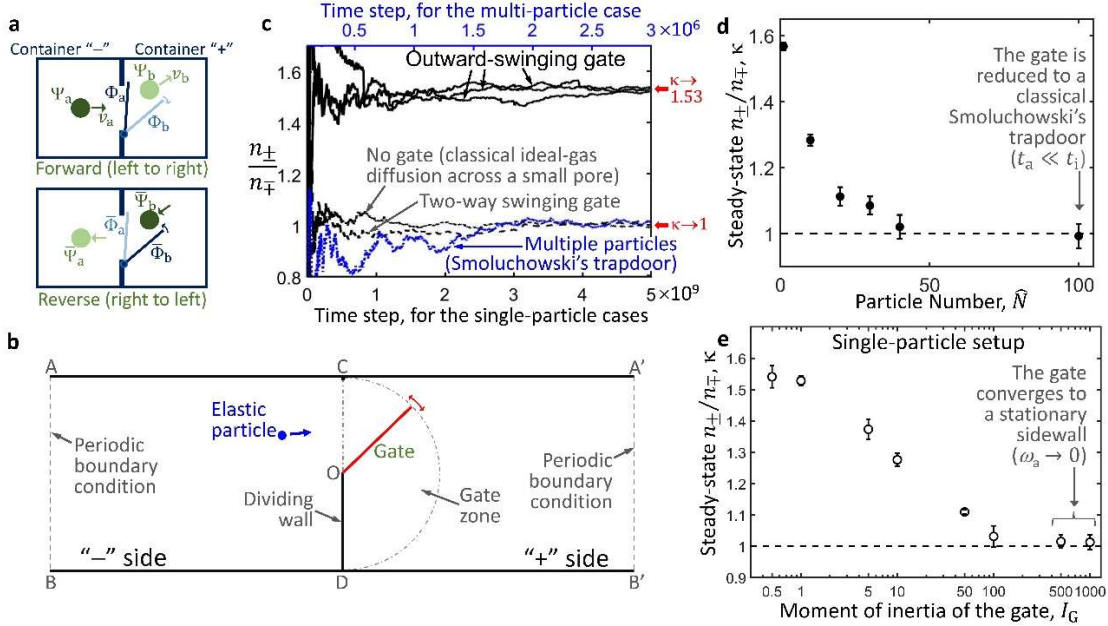

**Figure 2 (a)** Schematic of the particle crossing process. **(b)** The Monte Carlo (MC) simulation of a billiard-like particle interacting with an outward-swinging gate. **(c)** The MC simulation result of the time profiles of $n_\pm/n_\mp$. The moment of inertia of the gate ($I_G$) is 1. The steady-state $n_\pm/n_\mp$ indicates the overall crossing ratio of the gate, $\kappa$. The three dotted curves are for the reference cases. The upper ruler of the horizontal axis is for the multi-particle case (the blue dotted curve); the lower ruler of the horizontal axis is for the single-particle cases (all the black curves). **(d)** The steady-state $n_\pm/n_\mp$ (i.e., $\kappa$) as a function of the particle number ($\widehat{N}$). With multiple particles, the gate is reduced to a chaotic trapdoor, so that $\kappa$ decreases to 1. **(e)** The steady-state $n_\pm/n_\mp$ (i.e., $\kappa$) as a function of the moment of inertia of the gate ($I_G$), for the single-particle setup. As $I_G$ becomes larger, the gate is increasingly heavy and converges to a stationary sidewall, so that $\kappa$ decreases to 1.

In Figure 1(a), the two gas containers are large, in which the gas molecular motion is chaotic. The characteristic duration of the particle-gate interaction event ($t_i$) is much shorter than the characteristic time for an outgoing particle to come back to the gate ($t_r$), where $t_i$ accounts for the time for the particle to enter and exit the gate zone, as well as the time for the gate to return to equilibrium after the particle-gate collision. In such a system, the microstate of the incident particle ($\Psi_a$) is uncorrelated with the microstate of the gate ($\Phi_a$). If the particle distribution is uniform, the ensembles of the incident particles from right to left and from left to right are symmetric; the characteristic interval between incident particles ($t_a$) is the same in containers "−" and "+".

Like the nonequilibrium steady states of many other nonchaotic or nonergodic systems,[2-7] the overall nonuniformity in Figure 1(a) is consistent with the microscopic reversibility.[9] For each set of $\Psi_a$ and $\Phi_a$ in the forward process in Figure 2(a), there are a set of $\overline{\Psi}_b$ and $\overline{\Phi}_b$ of the reverse process. In either Hamiltonian dynamics or stochastic thermodynamics, regardless of ergodicity and chaoticity, time reversibility ensures microscopic reversibility:[10,11]

$$\{\Psi_b\Phi_b|\Psi_a\Phi_a\} = \{\overline{\Psi}_a\overline{\Phi}_a|\overline{\Psi}_b\overline{\Phi}_b\} \tag{1}$$

where $\{\bullet|\blacklozenge\}$ is the conditional probability of $\bullet$ given $\blacklozenge$. In a chaotic system, Equation (1) indicates that the probability of any process is symmetric in forward and reverse directions,[10] so that for Smoluchowski's trapdoor, $\kappa = 1$.[12,13]

In Figure 1(a), however, the gate is locally nonchaotic, specifically $t_a \gg t_i$. Under this condition, the particle-gate interaction events are independent of each other. Since only a subset of the incident particles can cross the gate ($\Psi_a\Phi_a$ in the forward process, $\overline{\Psi}_b\overline{\Phi}_b$ in the reverse process) and the others are blocked, the overall crossing ratio of the gate should be calculated as

$$\kappa = \frac{\delta_\pm}{\delta_\mp} = \frac{\int \{\Psi_b\Phi_b|\Psi_a\Phi_a\}\cdot\{\Psi_a\}\cdot\{\Phi_a\}d\Gamma}{\int \{\overline{\Psi}_a\overline{\Phi}_a|\overline{\Psi}_b\overline{\Phi}_b\}\cdot\{\overline{\Psi}_b\}\cdot\{\overline{\Phi}_b\}d\Gamma} \tag{2}$$

where $\{\bullet\}$ is the probability of microstate $\bullet$, and $\Gamma$ indicates the phase space. In the current system, $\{\overline{\Psi}_b\} = \{\Psi_b\}$, and $\{\overline{\Phi}_b\} = \{\Phi_b\}$.

Without extensive particle collision at the gate, in Equation (2), there is no mechanism for the system to reach thermodynamic equilibrium. Firstly, in the forward process from $\Psi_a$ to $\Psi_b$, upon the nonchaotic particle-gate interaction, $v_b$ is determined by the following parameters: $v_a$, the length ($L_G$) and the moment of inertia ($I_G$) of the gate, the angular velocity ($\omega_a$) and the swinging angle ($\phi_a$) of the gate at $\Psi_a$, the collision location on the gate ($L_c$), the particle mass ($m_P$) and the particle size ($D_P$), and the incident angle of the particle ($\psi_a$). Based on dimensional analysis, we have $\frac{v_b}{v_a} = \hat{f}\left(\frac{\omega_a L_G}{v_a}, \frac{3I_G/L_G^2}{m_P}, \frac{L_c}{L_G}, \frac{D_P}{L_G}, \phi_a, \psi_a\right)$, where $\hat{f}$ represents a certain function. When the thermal motion of the gate is significant (i.e., $\omega_a \neq 0$), $v_b$ is nonlinear to $v_a$, so that $v_a$ and $v_b$ do not have the same probability distribution; i.e., $\{\Psi_b\} \neq \{\Psi_a\}$, which leads to $\{\overline{\Psi}_b\} \neq \{\Psi_a\}$. That is, although the ensemble of incident particle is symmetric in both directions, the probability of crossing/blocking can be asymmetric.

Secondly, if there is an attraction force ($F_G$) between the gate and the "door stopper" on the dividing wall, the gate would be self-closing. In other words, $\Phi_a$ tends to be a closed configuration. Notice that when $D_P/L_G$ is nontrivial, $\Phi_b$ must be an open configuration, so is $\overline{\Phi}_b$. With $F_G$, because the closed gate configuration is energetically favorable but the open configuration is energetically unfavorable, $\{\overline{\Phi}_b\}$ is unequal to $\{\Phi_a\}$.

Consequently, since the probability distributions of $\overline{\Psi}_b$ and $\overline{\Phi}_b$ mismatch with those of $\Psi_a$ and $\Phi_a$, as the system parameters ($F_G$, $I_G$, $L_G$, $m_P$, $D_P$) may be arbitrarily chosen, Equation (2) suggests that in general, $\kappa \neq 1$.

In Section A1.i in Appendix, by directly solving the governing equations of particle-gate collision (for $F_G = 0$), we calculate the probability distributions of the crossing cases for the velocity of the particle, the angular coordinate of the particle, and the angular velocity of the gate. The result confirms that when $I_G$ is not too large, $\{\Psi_a\}\{\Phi_a\} \neq \{\overline{\Psi}_b\}\{\overline{\Phi}_b\}$.

In Figure 1(a), while in the interior of the containers the particle behavior is regular, the gate is a nonchaotic component. It changes the boundary condition. These factors are not considered in the classical Boltzmann equation and the H-theorem.[1]

2.2 Monte Carlo simulation

Figure 2(b) shows a Monte Carlo (MC) simulation of a billiard-like particle. The details are given in Section A2 in Appendix; the computer program is available at [14]. The particle randomly moves in a container. The container is separated into two sections by a dividing wall. The upper and the lower container walls and the dividing wall are diffusive. The left and the right borders are open, and use periodic boundary condition. The gate is a specular line.

Its swinging motion is limited to the "+" side by a "door stopper" at point C. We compare the total times the particle crosses the gate zone from right to left ($n_\mp$) and from left to right ($n_\pm$). The steady-state $n_\pm/n_\mp$ ratio indicates the overall crossing ratio of the gate, $\kappa$.

Figure 2(c) gives the numerical result of the time profiles of $n_\pm/n_\mp$, for $I_G = 1$. The three dotted lines are the reference curves. In one reference case, the gate is removed, so that the opening in the dividing wall (OC) remains unblocked; it represents the classical process of ideal-gas diffusion across a small pore, as discussed by Pauli.[15] In another reference case, the gate is not obstructed by point C, so that it can freely swing at both sides of the dividing wall. In the third reference case, the gate is outward-swinging, while there are 100 particles in the container; it is similar to the classical Smoluchowski's trapdoor,[12,13] for which the gate motion is chaotic (i.e., $t_a \ll t_i$). All the reference curves converge to 1, as they should. It confirms that the average particle arrival rates are equal at the two sides, and the local nonchaociticy is critical. The solid curves are for three randomized simulations of the outward-swinging gate with a single particle, for which the gate motion is nonchaotic (i.e., $t_a \gg t_i$). They use the same parameter setting as the single-particle reference case. For all of them, the steady-state $n_\pm/n_\mp \rightarrow 1.53$, suggesting that the crossing ratio is asymmetric.

In Figure 2(d), the particle number ($\hat{N}$) is varied, with everything else being the same as the solid curves in Figure 2(c). As more particles are in the system, the average arrival time ($t_a$) decreases. When $\hat{N}$ is large, $t_a \ll t_i$, i.e., the local nonchaoticity is lost. Under this condition, the system is reduced to the classical chaotic case of Smoluchowski's trapdoor, so that the steady-state $n_\pm/n_\mp$ decreases to 1, consistent with the study in [12,13].

In Figure 2(e), the gate mass is varied, with everything else being the same as the solid curves in Figure 2(c). The trend is clear that when $I_G$ increases, the steady-state $n_\pm/n_\mp$ is reduced to 1, as the gate converges to a stationary sidewall. It is consistent with the numerical result in Section A1.i in Appendix, as well as the previous dimensional analysis that $v_b$ is nonlinear to $v_a$ only when $\omega_a \neq 0$. When $I_G$ is relatively small, the steady-state $n_\pm/n_\mp$ (i.e., $\kappa$) is much greater than 1.

In Section A1.ii in Appendix, we solve the particle-gate collision equations to directly assess the nominal crossing ratio of the gate ($\kappa_o$). The computed $I_G - \kappa_o$ relationship is compatible with Figure 2(e). Moreover, all the simulation cases converged within ~$3\times10^9$ time steps; with the same parameter setting, all the randomized simulations had similar stead-state $\kappa$; since at the diffusive walls the reflected particle/gate velocity was random, the steady-state $\kappa$ had little history dependence.

In this MC simulation and Section A1.ii in Appendix, there is no long-range force on the gate; that is, $F_G = 0$. When $I_G$ is small, the unbalanced crossing ratio ($\kappa \neq 1$) should be attributed to the microstate of the particle, i.e., $\{\overline{\Psi}_b\} \neq \{\Psi_a\}$. The effect of $F_G$ on $\{\Phi_a\}$ and $\{\overline{\Phi}_b\}$ will be demonstrated by the molecular dynamics (MD) simulation in Section 7 below.

The outward-swinging gate is a type of spontaneously nonequilibrium dimension discussed in [8]. The above analysis of microscopic reversibility accounts for the entire phase space ($\Gamma$). The effects of the microstates of the gate and the particle are both taken into consideration. The self-closing force is optional; i.e., $F_G$ can be 0. The multi-particle reference curve in Figure 2(c) shows that when $t_a \ll t_i$, Figure 2(a) is reduced to a classical trapdoor, in agreement with the previous study on the second law of thermodynamics.[12,13] In Section 4 below, we will discuss the consistency between the asymmetric crossing ratio ($\kappa \neq 1$) and the principle of maximum entropy.

In the future, a parameterized study needs to be performed to further examine the effects of $F_G$, $m_P$, and $D_P/L_G$. To analyze the time-average properties, the characteristics of small systems should be taken into account.[e.g., 16]

## 3. Difference from classical models

### 3.1 Maxwell's demon and the Szilárd engine

The molecular-sized outward-swinging gate is not Maxwell's demon or its variant, e.g., the Szilárd engine.[17] When a Maxwell's demon guards an opening, its microstate depends on the microstate of the incident particle ($\Psi_a$),[18] which involves the physical nature of information.[19-21] On the contrary, in Figure 1(a) and Figure 2(a), there is no active information gathering or processing. The microstate of the gate ($\Phi_a$) is uncorrelated with the microstate of the incident particle ($\Psi_a$). The working mechanism is associated with the incomplete thermalization and the probability distributions of microstates (Equation 2). The gas containers in Figure 1(a) are large and chaotic.

3.2 Feynman's ratchet and Smoluchowski's trapdoor

Figure 1(a) has fundamental difference from Feynman's ratchet[22] and Smoluchowski's trapdoor.[12,13] These classical systems do not have nonchaotic components. Upon reaching thermodynamic equilibrium, they cannot spontaneously deviate from it, since the thermal motions of all the parts are balanced. For instance, in Feynman's model, the probability for the ratchet to overcome the energy barrier of the pawl ($\Delta E_p$) is dominated by $e^{-\beta \cdot \Delta E_p}$, where $\beta = 1/k_B T$, $k_B$ is the Boltzmann constant, and $T$ is temperature. The same Boltzmann factor also rules the Brownian movement of the vanes.

Conversely, in Figure 1(a), the local nonchaoticity of the gate motion is essential. It allows the system to reach a nonequilibrium steady state. For example, consider a self-closing gate with $F_G$. In the forward process, under the condition of local nonchaoticity ($t_a \gg t_i$), the probability for the gate to be pushed open by a gas particle is governed by $e^{-\beta(\Delta E_G - K_P)}$, where $\Delta E_G$ is the energy barrier caused by $F_G$, and $K_P$ is the kinetic energy from the incident particle. In the reverse process, the probability for the gate to spontaneously open is governed by $e^{-\beta \cdot \Delta E_G}$. Hence, when the gate is closed, the probabilities of crossing are unequal in the two directions. If the gate is at an open configuration as a particle arrives, the particle may pass through it without collision; such "leakage" events reduce but do not eliminate the overall asymmetry in $\kappa$. In comparison, for Smoluchowski's trapdoor, as the particle arrival rate is relatively high and the local nonchaoticity condition is broken (i.e., $t_a \ll t_i$), the mechanism of $e^{-\beta(\Delta E_G - K_P)}$ is irrelevant.[12,13]

Unlike Feynman's ratchet, the gate in Figure 1(a) does not rely on the fluctuation of any potential field (e.g., chemical potential), nor does it manipulate any potential field by changing pressure or volume. The gate is placed at the boundary of the large gas containers, and may be alternately exposed and covered by a frictionless sliding door. Thus, the system can shift between the nonequilibrium steady state and the equilibrium state.

3.3 Osmotic pressure

In Figure 1(a), the pressure difference across the gate ($\Delta P$) is not an osmotic pressure. An osmosis membrane is symmetric. It selectively obstructs particles according to their sizes. In osmosis, the crossing ratio for each type of particles is 1; the interaction among particles is not essential; for each crossing event, the probability distributions of the incident particle microstate and the outgoing particle microstate are the same; to measure $\Delta P$, the pressure sensors in the two containers must be exposed to different gaseous/liquid compositions; most importantly, the system cannot spontaneously deviate from thermodynamic equilibrium.[23]

4. Nonequilibrium maximum of entropy

In this section, we show that $\kappa \neq 1$ is compatible with the principle of maximum entropy, while $S$ may decrease in an isolated system.

As a first-order analysis, consider Figure 1(a) as a classical system with discrete microstates. For a canonical ensemble, the nonchaotic particle-gate interaction imposes a set of constraints on the probability of system microstates:

$$\frac{\rho_m}{\rho_n} = \kappa_{mn} \qquad (3)$$

where $\rho$ indicates probability, subscripts $m$ and $n$ indicate system microstates ($m = 1,2,3 \ldots$ and $n = m+1, m+2 \ldots$), $\kappa_{mn} = \kappa^{N_{mn}}$ is the probability ratio, and $N_{mn}$ is the excess number of gas molecules in container "+". That is, compared to microstate $n$, if microstate $m$ has $N_{mn}$ more gas molecules in container "+", $\rho_m$ would be different from $\rho_n$ by a factor of $\kappa^{N_{mn}}$. It can be viewed from the perspective of the equivalent potential difference ($\Delta\hat{E}$). In terms of the distribution of the particles, the gate in the locally nonchaotic system (Figure 1a) has the same effect as $\Delta\hat{E} = -k_B T \cdot \ln\kappa$ between containers "−" and "+" in a fully chaotic system; for the latter, it can be seen that $\frac{\rho_m}{\rho_n} = \exp\left(-\frac{N_{mn}\Delta\hat{E}}{k_B T}\right) = \kappa^{N_{mn}}$. In Figure 1(a), if in addition to the gate, there is also a potential difference ($\Delta E$) between the two containers, $\kappa_{mn}$ should be modified as $\tilde{\kappa}^{N_{mn}}$, where $\tilde{\kappa} = \kappa \cdot e^{-\beta \cdot \Delta E}$.

At thermodynamic equilibrium, entropy $S = -k_B \sum_i \rho_i \ln\rho_i$ reaches the maximum value ($S_{eq}$) with two constraints on $\rho_i$, $\sum_i \rho_i = 1$ and $\sum_i \rho_i E_i = U$,[1] where $\Sigma$ indicates summation for microstates ($i = 1,2,3 ...$), $E_i$ and $\rho_i$ are respectively the energy and the probability of the $i$-th microstate, and $U$ is the internal energy. With Equation (3), the Lagrangian becomes $\mathcal{L} = -k_B \sum_i \rho_i \ln\rho_i + \check{\lambda}(\sum_i \rho_i - 1) + \hat{\lambda}(\sum_i \rho_i E_i - U) + \sum_{m,n}[\lambda_{mn}(\rho_m - \kappa_{mn}\rho_n)]$, where $\check{\lambda}$, $\hat{\lambda}$, and $\lambda_{mn}$ are the Lagrange multipliers. To maximize $S$,[1]

$$\frac{\partial \mathcal{L}}{\partial \rho_i} = 0 \tag{4}$$

leads to $\rho_i = \exp\left[\frac{1}{k_B}(\varsigma_0 + \varsigma_i)\right]$, where $\varsigma_0 = -k_B + \check{\lambda} + \hat{\lambda}E_i$ and $\varsigma_i = \sum_{n>i} \lambda_{in} - \sum_{m<i} \kappa_{mi}\lambda_{mi}$. In accordance with $\sum_i \rho_i = 1$, we define $Z^* = \sum_i \exp\left[\frac{1}{k_B}(\hat{\lambda}U + \varsigma_i)\right] = \exp\left[-\frac{1}{k_B}(-k_B + \check{\lambda})\right]$ as the nonequilibrium partition function, which gives $\check{\lambda} = k_B \ln Z^* + k_B$. Thus, $\rho_i = \frac{1}{Z^*}\exp\left[\frac{1}{k_B}(\hat{\lambda}U + \varsigma_i)\right]$ and $S = -k_B \sum_i \rho_i \ln\rho_i = -\hat{\lambda}U - \sum_i \rho_i \varsigma_i + k_B \ln Z^*$. Because $\frac{dS}{dU} = \frac{1}{T}$, $\hat{\lambda} = -\frac{1}{T}$. Substitution of the expression of $\rho_i$ into Equation (3) suggests that $\varsigma_i = k_B N_i^+ \ln\kappa$, where $N_i^+$ is the number of gas molecules in container "+" of the $i$-th microstate. Consequently,

$$Z^* = e^{-\beta U} \sum_i \kappa^{N_i^+} \tag{5}$$

$$\rho_i = \frac{1}{\tilde{Z}^*} \kappa^{N_i^+} \tag{6}$$

$$S = \frac{U}{T} + k_B \ln Z^* - \frac{k_B \ln\kappa}{\tilde{Z}^*} \sum_i N_i^+ \kappa^{N_i^+} \tag{7}$$

where $\tilde{Z}^* = Z^* e^{\beta U} = \sum_i \kappa^{N_i^+}$. If different microstates may have different $E_i$, through a similar derivation procedure, we have $Z^* = \sum_i e^{-\beta E_i} \kappa^{N_i^+}$, $\rho_i = \frac{1}{Z^*} e^{-\beta E_i} \kappa^{N_i^+}$, and $S = \frac{U}{T} + k_B \ln Z^* - \frac{k_B \ln \kappa}{Z^*} \sum_i e^{-\beta E_i} N_i^+ \kappa^{N_i^+}$.

According to Equation (7), when $\kappa \neq 1$, as $S$ is maximized by Equation (4), it reaches a nonequilibrium maximum ($S_{ne}$), which is less than $S_{eq}$. At thermodynamic equilibrium, the gas pressures in containers "–" and "+" are the same, denoted by $P_0$. With the molecular-sized outward-swinging gate, the steady-state pressure ratio $P_+/P_- = \kappa$, where $P_-$ and $P_+$ are the gas pressures in containers "–" and "+", respectively. Because $P_- + P_+ = 2P_0$, we have $P_- = \frac{2}{1+\kappa} P_0$ and $P_+ = \frac{2\kappa}{1+\kappa} P_0$. Therefore, $N_- = \frac{2}{1+\kappa} N$ and $N_+ = \frac{2\kappa}{1+\kappa} N$, where $N = \frac{P_0 V_0}{k_B T}$, $V_0$ is the container volume, and $N_-$ and $N_+$ are the numbers of gas molecules in containers "–" and "+", respectively. From the equation of entropy of ideal gas,[1] the decrease in entropy ($\Delta S = S_{ne} - S_{eq}$) can be calculated as

$$\Delta S = \left( N_- k_B \ln \frac{eV_0}{N_-} + N_+ k_B \ln \frac{eV_0}{N_+} \right) - \left( 2 \cdot N k_B \ln \frac{eV_0}{N} \right) = -2N k_B \tilde{f} \qquad (8)$$

where $\tilde{f} = \ln \left( \frac{2}{\kappa+1} \kappa^{\frac{\kappa}{\kappa+1}} \right)$. With a constant $U$, the associated increase in Helmholtz free energy is $\Delta F = -T\Delta S = 2N k_B T \tilde{f}$. If $d\kappa = \kappa - 1$ is small, $\Delta S \approx -\frac{N k_B}{4} d\kappa^2$ and $\Delta F \approx \frac{N k_B T}{4} d\kappa^2$.

With the molecular-sized outward-swinging gate, $S = S_{ne}$; if the containers are connected through a regular open channel, $S = S_{eq}$. Hence, as the connection between the two containers is changed from a regular channel to a molecular-sized outward-swinging gate, $S$ decreases by $\Delta S$ from $S_{eq}$ (the global maximum) to $S_{ne}$ (a local maximum), without an energetic penalty. Before and after the transition, $S$ remains maximized, since Equation (4) is always satisfied. The decrease of $S$ is caused by the reduction in the maximum possible entropy of steady state ($S_Q$), as the gate influences the boundary condition of the gas containers.

The internal gas diffusion does not cause an overall heat exchange with the environment, and the internal energy is constant. The result of the above analysis is also applicable to a microcanonical ensemble. The same Equation (8) can be used to calculate $\Delta S$ caused by the diffusive gas transfer either in an adiabatic process (e.g., Figure 1a) or in an isothermal process (e.g., the experiment discussed below).

## 5. Experimental design

One method to experimentally investigate molecular-sized gates is to surface-graft molecular chains at nanopore openings.[e.g., 24,25] As the carbon-carbon or carbon-nitrogen bonds rotate, an organic chain can be bent.[e.g., 26,27] The van der Waals attraction force between the substrate and the grafted chains may serve as $F_G$.

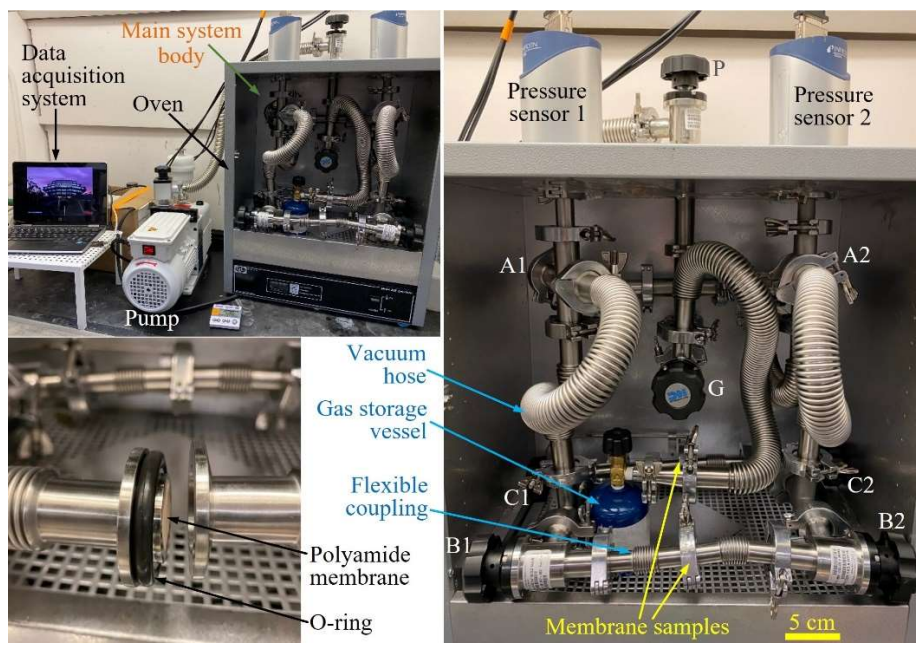

**Figure 3.** The gas-pressure measurement system (upper left), the main system body (right), and a polyamide membrane mounted on a compound o-ring (bottom left). The letters in the photo of the system body indicate the vacuum valves. The system body is placed in a QL model-30GCE box oven.

Figure 1(b) and Figure 3 show the experimental setup. The details of the testing procedure are given in Section A3 in Appendix. We used lauric aldehyde (LA) for the surface grafting. The molecular structure of LA is depicted in Figure 4(a). It has 12 carbon atoms, with the molecular mass ($m_c$) ~184 and the contour length ~14 Å. It is somewhat similar to the bendable organic chains studied by Kim et al.,[28] but has different end groups. One end is a methyl group (–CH$_3$), which is nonpolar;[29] the other end is an aldehyde group (–CHO), which is reactive to amide linkage,[30] as illustrated in Figure 4(c).

Grafting of dodecyl chains was performed on the front surface of a 10 μm-thick nanoporous polyamide membrane, as shown in Figures 4(d,e); the back surface of the membrane was untreated. The nanopore size was below 1 nm. The grafted side was toward container 2. The gas phase was pentafluoroiodoethane (C$_2$F$_5$I) (Figure 4b). Its molecular mass ($m_g$) is ~246 and the molecular size is ~6 Å. The gas pressures in container 1 ($P_1$) and container 2 ($P_2$) were continuously monitored by two pressure sensors, respectively. There were three valved channels between the two gas containers: a regular open hose between valves A1 and A2, a one-sidedly surface-treated membrane between valves B1 and B2, and an untreated membrane (or a nonpermeable solid film) between valves C1 and C2.

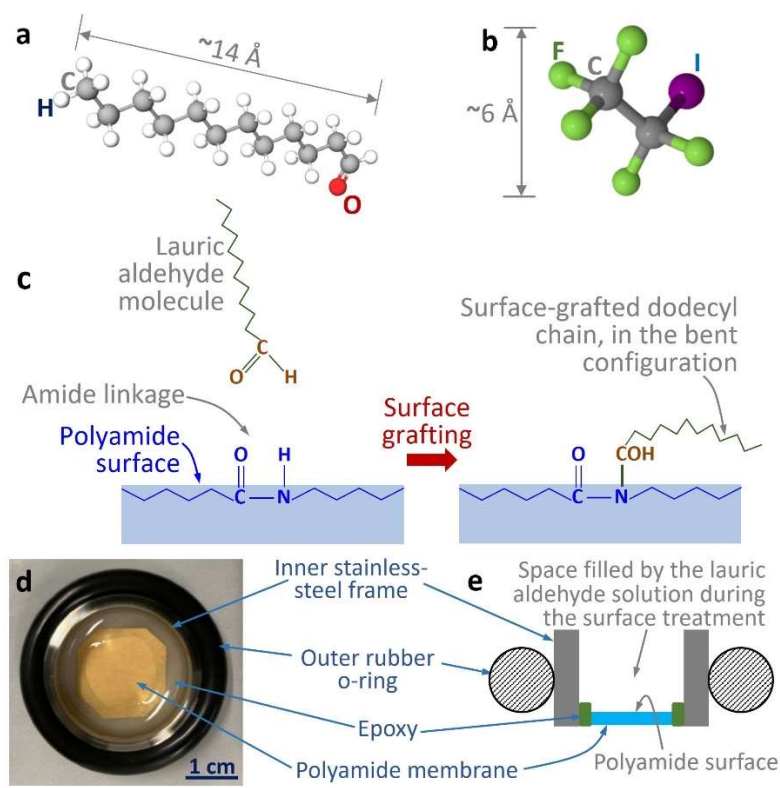

**Figure 4.** Schematics of **(a)** a lauric aldehyde (LA) molecule and **(b)** a pentafluoroiodoethane (C$_2$F$_5$I) gas molecule. **(c)** The aldehyde group (–COH) can react with an amide linkage. Thus, dodecyl chains can be grafted on a polyamide surface. **(d)** Top view and **(e)** a schematic of the cross section of a polyamide membrane mounted on a compound o-ring.

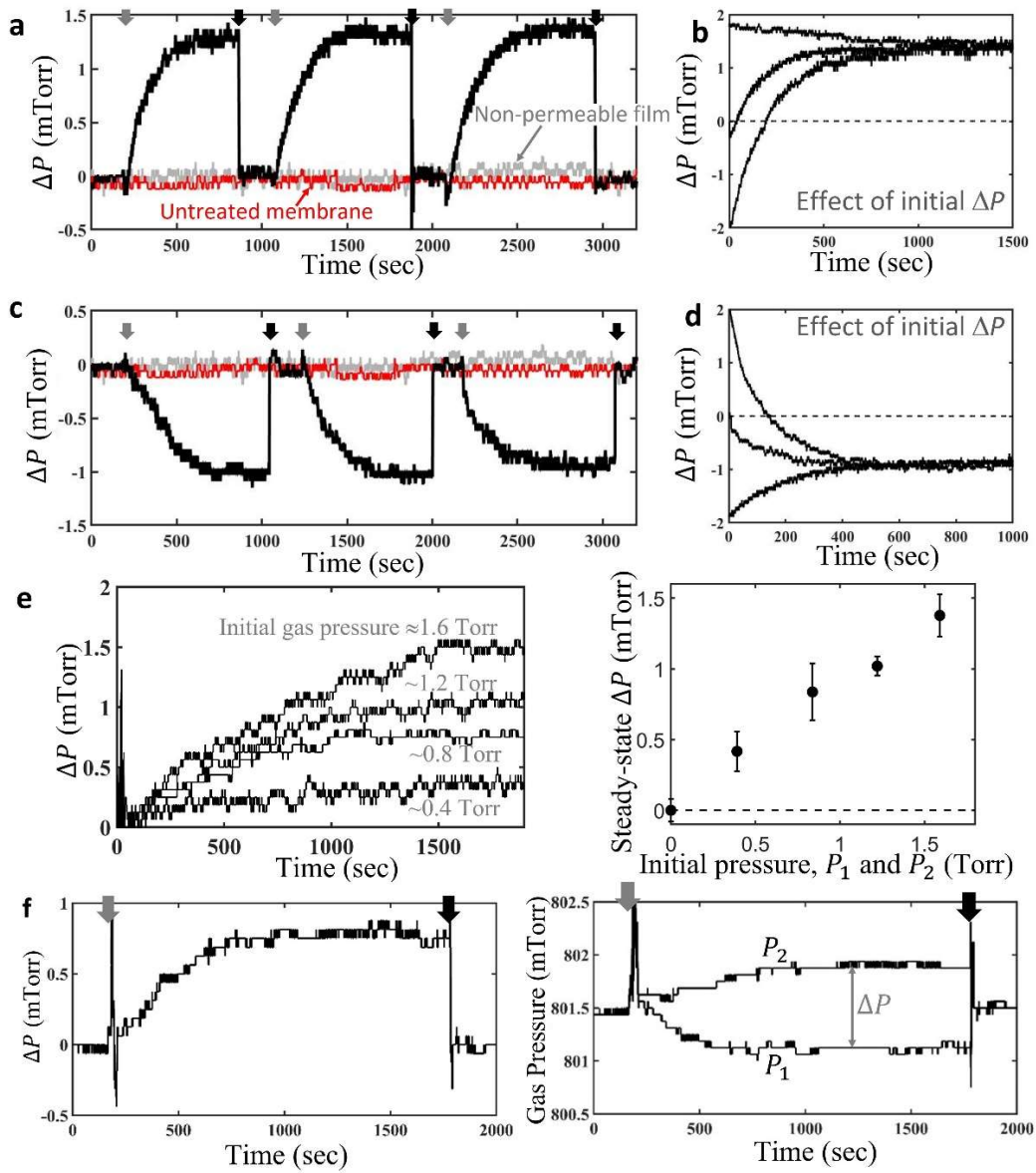

**Figure 5. (a)** Time profiles of the pressure difference ($\Delta P = P_2 - P_1$). The surface-grafted side faces container 2. The black curves are for the one-sidedly surface-grafted membrane; the red curve is for the untreated membrane; the gray curve is for the non-permeable solid film. The two valves across the membrane/film remain open; valves A1 and A2 are closed and opened repeatedly; all the other valves remain shut. The gray arrows indicate that valves A1 and A2 are closed; the black arrows indicate that valves A1 and A2 are reopened. **(b)** The initial $\Delta P$ has no influence on the steady-state $\Delta P$: $\Delta P$ is first adjusted to about -2, 0, or 2 mTorr; with valves B1 and B2 being open and all the other valves being closed, $\Delta P$ eventually converges to the same steady state as in Figure 5a. The surface-grafted side faces container 2. **(c)** Similar to Figure 5a, while the surface-grafted side faces container 1. The red and the gray curves are the same as in (a). **(d)** Similar to Figure 5b, while the surface-grafted side faces container 1. **(e)** When the initial gas pressure ($P_1$ and $P_2$) changes from ~0.4 Torr to ~1.6 Torr, the steady-state $\Delta P$ increases nearly proportionally. Left: typical time profiles of $\Delta P$; right: the steady-state $\Delta P$ as a function of the initial gas pressure. **(f)** Associated with the development of $\Delta P$ (left), $P_2$ increases by ~$\Delta P/2$ and $P_1$ decreases by ~$\Delta P/2$ (right). The surface-treated side faces container 2.

The tests were conducted at ambient temperature ~22 °C. Valves A1, A2, B1, and B2 were initially open, and valves P, G, C1, and C2 remained closed. The initial $P_1$ and $P_2$ were ~0.8 Torr. At such a pressure, the average spacing among gas molecules was ~30 nm, much larger than the chain length. Since the effective permeability of the open hose between valves A1 and A2 was higher than the membrane permeability by many orders of magnitude, the initial pressure difference measured by the two pressure sensors ($\Delta P = P_2 - P_1$) was zero.

## 6. Experimental results and discussion

Initially, $\Delta P \approx 0$. We closed valves A1 and A2, leaving only valves B1 and B2 open. Figure 5(a) shows that spontaneously, a pressure difference was developed across the one-sidedly surface-treated membrane. In ~9 min, $\Delta P$ reached ~1.2 mTorr. It was positive, indicating that $P_2 > P_1$; that is, gas flew from the low-pressure side (container 1) to the high-pressure side (container 2), until the steady state was reached. The effective gas permeability, $\dot{J}/(P_1 - P_2)$, was negative, where $\dot{J}$ is the average gas flow rate.[31] After $\Delta P$ has stabilized for ~5 min, valves A1 and A2 were reopened. The pressure difference instantaneously decreased to zero. As valves A1 and A2 were closed and opened again, the increase and decrease of $\Delta P$ were repeatedly observed. If the membrane was flipped and the surface-grafted side faced container 1, a similar $\Delta P$ profile was measured, except that $\Delta P = P_2 - P_1$ became negative (Figure 5c). The slight difference between the steady-state $\Delta P$ in Figure 5a and Figure 5c was due to data scatter. The steady-state $\Delta P$ was stable for more than 12 h (Figure 6).

The container volume ($V_0$) was about 710 cm³. With $P_1$ and $P_2$ being ~0.8 Torr, according to the ideal-gas law,[1] the amount of gas in each container was ~3.1×10⁻⁵ mol. Upon reaching the steady-state $\Delta P$, approximately 2.3×10⁻⁸ mol gas has transferred across the membrane. Equation (8) suggests that $\Delta S \approx -1.6 \times 10^{-10}$ J/K and $\Delta F \approx 4.8 \times 10^{-8}$ J.

If valves C1 and C2 were open and all the other valves were shut, $\Delta P$ was nearly zero (the red curve in Figure 5a). That is, the gas pressure across an untreated symmetric membrane was balanced, as it should be. If we replaced the untreated membrane between valves C1 and C2 by a non-permeable solid film, the change in $\Delta P$ was also trivial over time (the gray curve in Figure 5a). Only when the two containers were separated by a one-sidedly surface-treated membrane, could $P_1$ and $P_2$ be different.

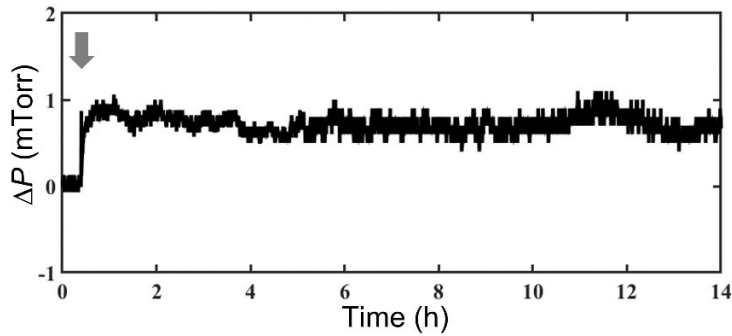

**Figure 6.** Long-time measurement of the gas pressure difference ($\Delta P$). The surface treatment and the testing procedures were similar with those of Figure 5(a). The gray arrow indicates that valves A1 and A2 are closed.

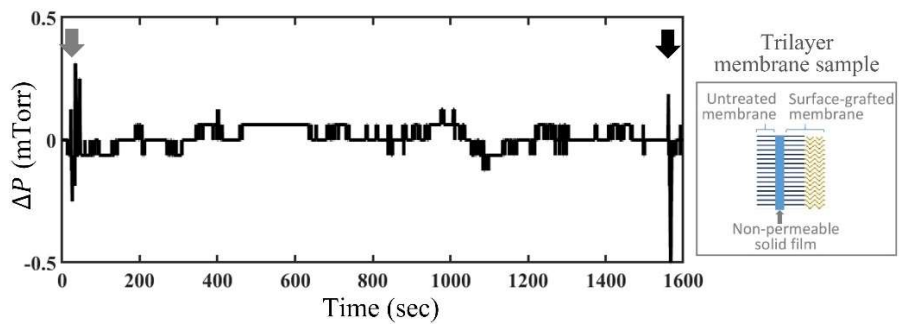

**Figure 7.** The pressure difference ($\Delta P$) across a trilayer membrane remains near zero. The trilayer sample is depicted by the inset at the right-hand side. It has the same front and back surfaces as a one-sidedly surface-treated membrane, but the middle layer is nonpermeable and therefore, no gas transport can take place. The testing result confirms that the effect of the gas adsorption and desorption of the membrane is negligible. The gray arrow indicates that valves A1 and A2 are closed; the black arrow indicates that valves A1 and A2 are reopened.

The sign of the steady-state $\Delta P$ followed the membrane direction (Figures 5a,c). Figures 5(b,d) and Figure 5(e) show that the steady-state $\Delta P$ is independent of the initial $\Delta P$, and proportional to $P_1$ and $P_2$. Associated with the development of $\Delta P$, the gas pressure at the surface-grafted side increased by $\sim\Delta P/2$, and the gas pressure at the back side decreased by $\sim\Delta P/2$ (Figure 5f), indicating a mass transfer across the membrane. The pressure sensors were ~90 cm away from the membrane (Figure 3), ensuring that the measured $\Delta P$ was not a local phenomenon. The container was ~710 cm$^3$ in volume, in which the effect of gas adsorption and desorption of the 1.3-cm$^2$ membrane was negligible (Figure 7). If a dodecane layer was not chemically bonded to the membrane but physically adsorbed, no $\Delta P$ could be detected (Figure 8a); likewise, if the grafted side of the surface-treated membrane was physically attached to an untreated membrane, the steady-state $\Delta P$ remained similar (Figure 8b), demonstrating that $\Delta P$ must be attributed to the covalent bonding between the grafted chains and the membrane surface, i.e., the gate-like chain behavior. These observations suggest that the system steady state was dominated by the overall crossing ratio of the membrane ($\bar{\kappa}$), which determines the final $P_2/P_1$ ratio.

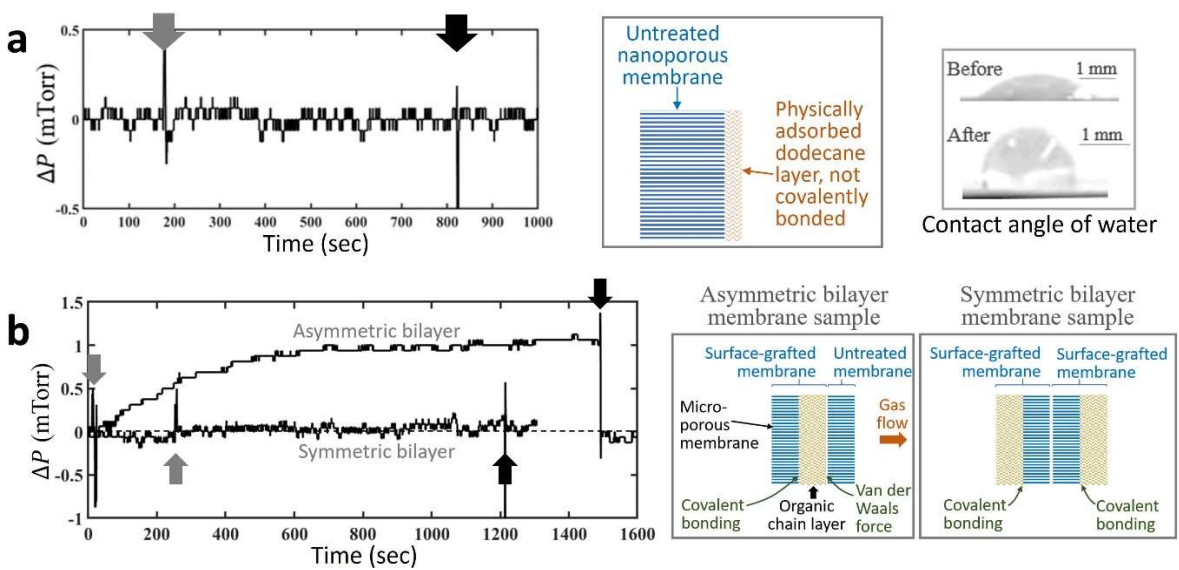

**Figure 8.** The pressure difference ($\Delta P$) was caused by the gate-like chain behavior. **(a)** A dodecane layer is not chemically grafted, but physically adsorbed on the membrane surface through solvent deposition, as depicted by the inset in the middle. No $\Delta P$ can be detected, showing that without the covalent bonding with the membrane surface, the organic chains do not drive the system away from thermodynamic equilibrium. **(b)** Across an asymmetric bilayer sample, the steady-state $\Delta P$ is similar with that of the one-sidedly surface-grafted membrane; across a symmetric bilayer sample, $\Delta P$ remains nearly zero. The gray arrows indicate that valves A1 and A2 are closed; the black arrows indicate that valves A1 and A2 are reopened.

## 7. Overall crossing ratio of the membrane

We used $C_2F_5I$ as the gas phase, because of its large $m_g$. In general, the gas particles must be sufficiently heavy to overcome the organic chains (see Section A3.x in Appendix).

If all the nanopore openings were perfectly grafted and all the dodecyl chains were perfect self-closing gates, regardless of the probability for a gas particle to push open a chain ($\chi$), the steady-state $P_2/P_1$ ratio, $\bar{\kappa}$, would be infinity. In the experiment, however, $\bar{\kappa}$ is finite. It should be attributed to the "leakage" of the open pores; i.e., not all the nanopores are obstructed by the organic chains. Firstly, a certain portion of the grafted chains are randomly in the straight configuration normal to the membrane surface (see Figure 9a). Secondly, while the contour chain length is ~1.4 nm, the average end-to-end distance is shorter, which may not be sufficient to cover the largest pores. Thirdly, to minimize the membrane deformation, the surface treatment temperature is relatively low and the treatment time is relatively short, and the amount of the LA solution is small; consequently, not all the amide linkages are end-capped.

With the percentage of the effectively covered pores being denoted by $\xi$, $\bar{\kappa}$ may be calculated as $1 + \frac{\xi\chi}{1-\xi}$. The rotational barrier of carbon-carbon bond, $E_C$, is on the scale of $10^4$ J/mol.[32] At room temperature, the associated Boltzmann factor is $\delta_0 = \exp\left(-\frac{E_C}{RT}\right) \approx 1.8\%$, where $R$ is the gas constant. When a gas particle impacts a chain, since the momentum along the chain backbone direction does not directly contribute to the chain rotation, as a first-order approximation, $\chi$ is $\delta_0^{3/2} \approx 0.24\%$. According to the experimental data in Figure 5(a), $\bar{\kappa} = P_2/P_1 \approx 1 + 0.16\%$. When ~40% of the nanopore openings are effectively gated (i.e., $\xi \approx 0.4$), $1 + \frac{\xi\chi}{1-\xi}$ is close to the measured $\bar{\kappa}$.

The above analysis of $\chi \approx 0.24\%$ is qualitatively in agreement with our molecular dynamics (MD) simulation, as detailed in Section A4 in Appendix. A carbon nanotube (CNT) was employed as an analogue to a micropore, at one end of which a dodecyl chain was covalently bonded. The gas particle was a mercury (Hg) atom, which collided with the dodecyl chain from either inside or outside of the CNT.

The MD simulation also gives the probability distribution of the tilting angle ($\theta_G$) of the equilibrium state of the grafted chain (Figure 9a). We randomly generated 100 chain configurations. Due to the van der Waals force between the CNT and the chain ($F_G$), the average $\theta_G$ is only 25.6°; that is, the chain tends to be self-closed, causing $\{\overline{\Phi}_b\} \neq \{\Phi_a\}$. Under this condition, when a gas particle (a mercury atom) moves upwards inside the CNT, it may push open the chain and cross the opening (Figure 4b); when the gas particle moves downwards to the CNT, it may push close the chain and be blocked (Figure 4c). Figure 9(d) shows the statistical result of the probability of crossing ($\delta_{cr}$) for the 100 random chain configurations. For all the initial gas-particle velocities ($v_0$) under investigation, $\delta_{cr}$ is always greater in the forward process than in the reverse process. Thus, the overall crossing ratio is asymmetric (i.e., $\kappa \neq 1$).

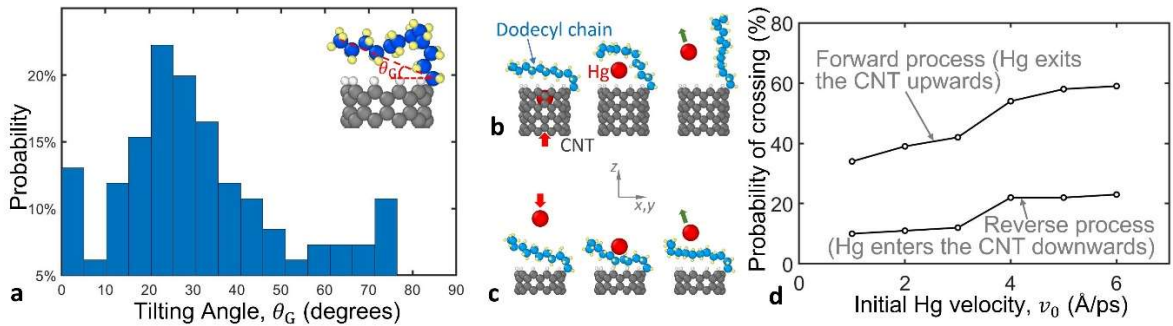

**Figure 9**. **(a)** Molecular dynamics (MD) simulation result of the probability distribution of the tilting angle ($\theta_G$), between the end surface of the carbon nanotube (CNT) and the end-to-end line of the covalently bonded dodecyl chain. The chain ends are defined as the centers of the first and the last carbon atoms. Altogether 100 equilibrium configurations were generated, among which 11 chains congested the CNT (i.e., $\theta_G < 0$) and are not included in this chart. **(b)** Side view of a mercury (Hg) atom exiting the CNT: before (left), during (middle), and after (right) the Hg-chain collision. The Hg atom moves upwards. **(c)** A Hg atom is blocked by the chain and cannot enter the CNT: before (left), during (middle), and after (right) the collision. The Hg atom initially moves downwards. **(d)** The probability of crossing ($\delta_{cr}$) as a function of the initial Hg velocity ($v_0$), for the forward and the reverse processes. The details of the MD simulation are given in Section A4 in Appendix.

There are other factors that also affect $\overline{\kappa}$. The membrane-chain attraction ($F_G$) increases the energy barrier of chain opening. The pore geometry, the collision mode, and the gas particle structure influence the dynamics of the gas-chain interaction. Reducing $E_C$ or raising $\xi$ may help to increase $\Delta P$. For instance, the rotational barrier of siloxane linkage is only ~8 meV,[33] so that $\chi$ of polysiloxane tends to be large. The recent study in molecular engineering may offer other low-barrier mechanisms of gate motion,[34,35] such as interlocked molecular rings. To enhance $\xi$, the membrane should be compatible with higher-temperature and longer-time surface treatment. Probably more importantly, the pore size and the pore shape must be uniform. In future study, the statistical fluctuation, the pore coverage ratio, and the pressure and temperature effects need to be examined in detail.

The experiment was performed at a constant temperature without thermal insulation, consistent with the canonical ensemble in the theoretical analysis in Section 4. As the internal energy of the gas phase remains unchanged, the diffusive gas transfer does not cause an overall heat exchange with the environment. Hence, the isothermal process is also adiabatic, corresponding to the isolated configuration in Figure 1(a). The asymmetric $\bar{\kappa}$ raises an interesting question: Whether useful work can be produced in a cycle by absorbing heat from a single thermal reservoir (see the discussion in Section A5 in Appendix).

## 8. Generalized form of the second law of thermodynamics

Conventionally, the second law of thermodynamics can be expressed in a number of equivalent forms.[1] The entropy statement claims that in an isolated system, entropy cannot decrease. The Kelvin-Planck statement claims that no useful work can be produced in a cycle from a single heat reservoir.

With arbitrary chaoticity and ergodicity, in accordance with the discussion in Section 4, the second law of thermodynamics may be generalized as follows: In an isolated system, entropy ($S$) cannot evolve away from the maximum possible value of steady state ($S_Q$), i.e.,

$$S \to S_Q \qquad (9)$$

In other words, in an isolated system, $|S - S_Q|$ can never increase; that is, entropy always has the tendency to converge toward $S_Q$, where $S_Q$ is determined by the principle of maximum entropy (e.g., Equation 4). This statement is based on the understanding that probability can be measured by entropy. It reflects the fundamental concept that a state of a higher probability is more probable to occur.

The general form (Equation 9) is broader than the classical entropy statement. They are equivalent to each other if $S_Q$ is equal to the maximum possible entropy of the system, i.e., the equilibrium maximum ($S_{eq}$). However, the general form allows for that $S_Q$ can be a nonequilibrium maximum ($S_{ne}$), which is less than $S_{eq}$. While $S$ always tends to be maximized (in the sense that the steady state is governed by Equation 4), for a nonchaotic or nonergodic system, a spontaneously nonequilibrium dimension may impose additional constraints on $\rho_i$. Thus, $S_Q$ can be reduced without an energetic penalty, e.g., from $S_{eq}$ to $S_{ne}$ when the boundary condition is changed by a molecular-sized outward-swinging gate. In an isolated system, if initially $S > S_Q$, entropy would decrease.

## 9. Concluding remarks

In the current research, we investigate the concept of molecular-sized outward-swinging gate, through theoretical analysis (Sections 2.1 and 4), Monte Carlo simulation (Section 2.2), directly solving the governing equations of particle-gate collision (Section A1.ii in Appendix), and experiment (Sections 5 and 6). All the results consistently indicate that under the condition of local nonchaoticity ($t_a \gg t_i$), the probabilities of particle crossing are unequal in the forward and the reverse directions.

The experiment was performed by using a nanoporous polyamide membrane one-sidedly surface-grafted with dodecyl chains. The membrane was placed in between two large containers filled with pentafluoroiodoethane gas. Remarkably, a pressure difference was repeatedly developed across the membrane, as the gas spontaneously flew from the low-pressure side to the high-pressure side. That is, the gas permeability of the membrane was asymmetric, and could be effectively negative.

The gate is a locally nonchaotic entropy barrier, which is a type of spontaneously nonequilibrium dimension [8]. It interrupts the probability distribution of the local microstates, and imposes additional constraints on the global microstates. Thus, entropy reaches a nonequilibrium maximum, less than the equilibrium maximum. It is compatible with microscopic reversibility, and is fundamentally different from Maxwell's demon, Feynman's ratchet, Smoluchowski's trapdoor, and osmosis. Essentially, the gate changes the boundary condition of the gas containers, so that the maximum possible entropy of steady state ($S_Q$) is reduced. Such a system can shift from the equilibrium state to the nonequilibrium steady state, causing an entropy decrease without an energetic penalty; e.g., entropy can decrease in an isolated system. It implies that useful work may be produced in a cycle by absorbing heat from a single thermal reservoir. To circumvent the contradiction to the conventional theory, the second law of thermodynamics may be generalized as Equation 9: In an isolated system, entropy cannot evolve away from $S_Q$.

## APPENDIX

**A1. Analyses of particle-gate interaction**

A1.i Probability distributions of microstates

Consider the two-dimensional system depicted in Figure 10(a). A billiard-like particle impacts a gate with the incident velocity of $\vec{v}_a$. The velocity of the outgoing particle is $\vec{v}_b$. The setup is scalable; an example of the unit system can be based on g/mole, Å, fs, and K. The gate is a rigid specular line, with the length ($L_G$) of 10; temperature ($T$) is set to 1000; the particle is a point mass of 1; the moment of inertia of the gate ($I_G$) varies in a broad range. The gate rotates freely around point O. The particle-gate collision is elastic and friction-free. No long-range force exists.

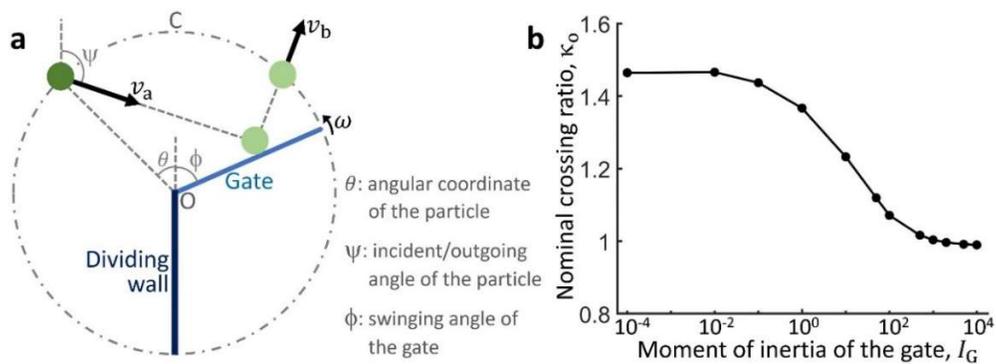

**Figure 10 (a)** Schematic of the particle-gate interaction. **(b)** The nominal crossing ratio of the gate ($\kappa_o$) calculated by directly solving the governing equations of the particle-gate collision. It is consistent with the result of the Monte Carlo simulation in Figure 2(e).

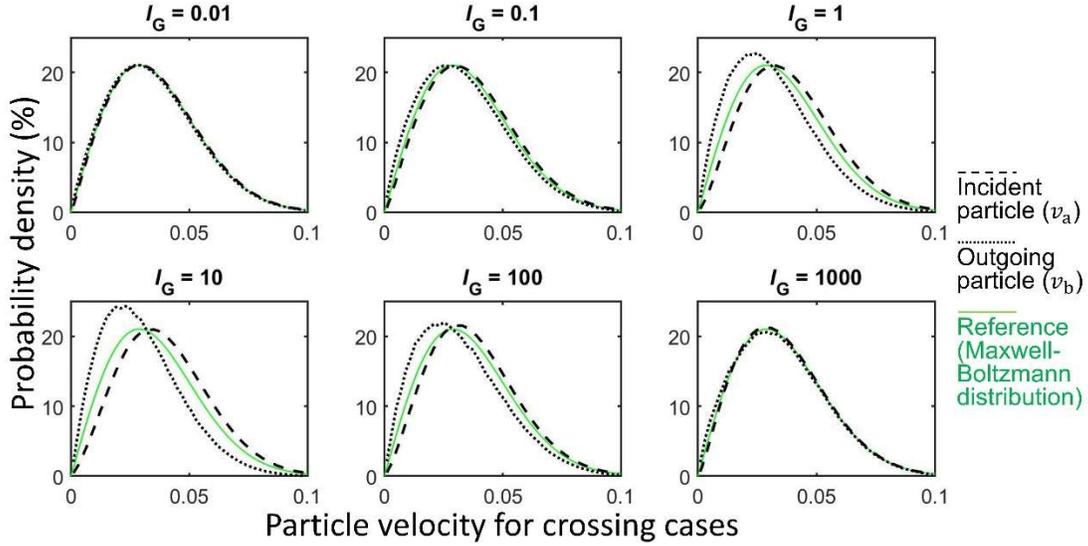
**Figure 11.** Probability density of the particle velocity for all the crossing cases.

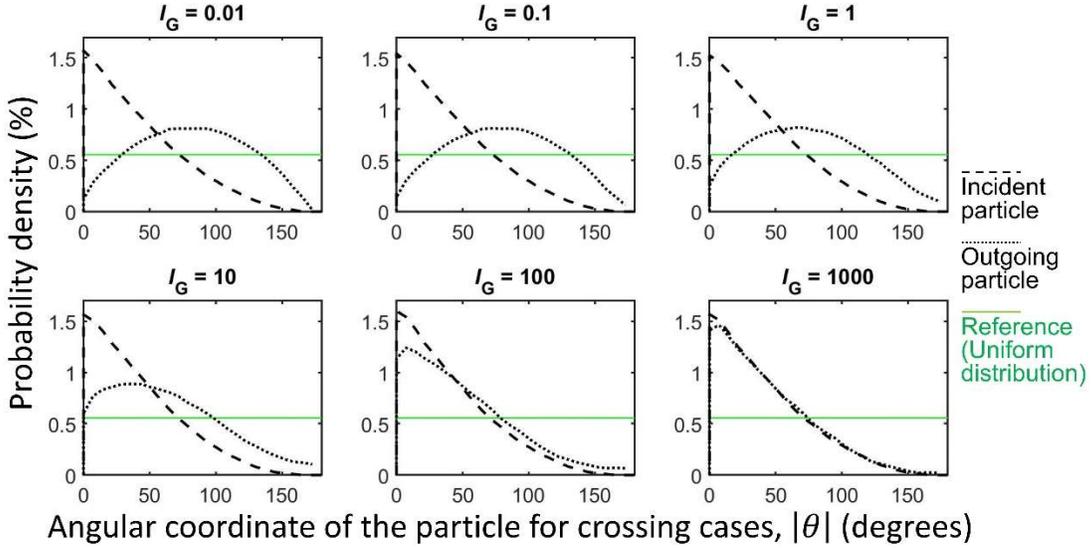
**Figure 12.** Spatial distribution of the particle for all the crossing cases.

The radius of the dash-dotted circle in Figure 10(a) is 10, the same as the gate size. Initially, the particle is placed on the left-hand side of the circle with given $v_a$, $\theta$, $\psi$, $\omega_a$, and $\phi$, where $\theta$ is the angular coordinate of the particle, $\psi$ is the direction of the particle velocity, $\omega_a$ is the angular velocity of the gate, and $\phi$ is the swinging angle of the gate. For difference calculation cases, $\theta$ varies from 0 to $\pi$, with the resolution of $\pi/50$; $\phi$ varies from 0 to $\pi$, with the resolution of $\pi/25$; $\psi$ varies from 0 to $\pi$, with the resolution of $\pi/50$; $v_a$ varies from 0 to 0.1, with the resolution of 0.002; $\omega_a$ varies from $-3\sqrt{k_B T/I_G}$ to $3\sqrt{k_B T/I_G}$, with the resolution of 1/50 of the full range.

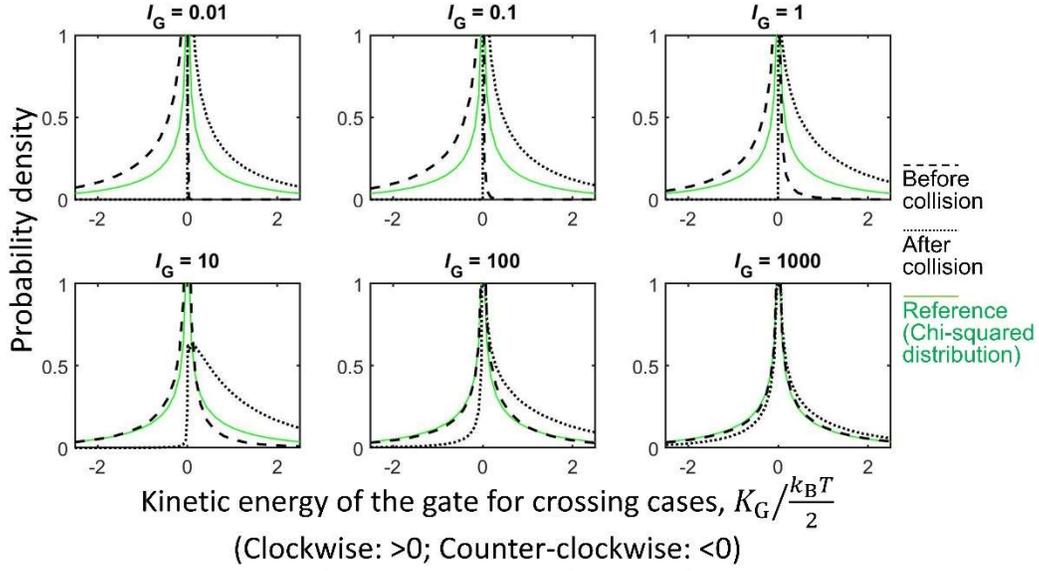

**Figure 13.** Distribute of the kinetic energy of the gate for all the crossing cases.

By solving the equations of energy conservation, conservation of angular momentum, and conservation of linear momentum along the gate line, for each case, after the particle-gate interaction, we obtain the particle velocity ($v_b$), the angular coordinate of the outgoing particle when it reaches the dash-dotted circle ($\theta$), and the angular velocity of the gate ($\omega_b$). The kinetic energy of the gate is $K_G = I_G \omega^2/2$, where $\omega$ is either $\omega_a$ or $\omega_b$. If the particle collides with the gate and can reach the right-hand side of the dash-dotted circle, the case is registered as crossing; otherwise, the case is dismissed.

It is assumed that $\theta$, $\psi$, and $\phi$ associated with the incident particle (i.e., microstates $\Psi_a$ and $\Phi_a$) are uniformly distributed; $v_a$ follows the two-dimensional Maxwell-Boltzmann distribution; $\omega_a$ follows the one-dimensional Maxwell-Boltzmann distribution. As suggested by Equation (2), only the crossing cases are essential to the calculation of the crossing ratio. For all the crossing cases, we compute the probability distributions of the particle velocity, the magnitude of $\theta$, and $K_G$, weighted by the distributions before particle-gate collision. The computer program is available at [14].

The results are shown in Figures 11-13, for various $I_G$. It can be seen that for the crossing cases, the distributions of these parameters are generally different from the overall distributions (the Maxwell-Boltzmann distribution for the particle velocity; the uniform distribution for $\theta$; the chi-squared distribution for $K_G$). More importantly, the distributions before collision (microstates $\Psi_a$ and $\Phi_a$) may not be the same as the distributions after collision (microstates $\Psi_b$ and $\Phi_b$), depending on $I_G$. When $I_G$ is large, the distributions of all the parameters are nearly symmetric for incident and outgoing particles, suggesting that $\{\Psi_a\} = \{\overline{\Psi}_b\}$ and $\{\Phi_a\} = \{\overline{\Phi}_b\}$. When $I_G$ is small, the distributions of the particle velocity and $K_G$ are nearly symmetric for incident and outgoing particles, yet the spatial distribution of the particle ($\theta$) is significantly different; thus, $\{\Phi_a\}$ tends to be the same as $\{\overline{\Phi}_b\}$, while $\{\Psi_a\}$ is different from $\{\overline{\Psi}_b\}$, so that $\{\Phi_a\} \cdot \{\Psi_a\} \neq \{\overline{\Phi}_b\} \cdot \{\overline{\Psi}_b\}$. When $I_G$ is in the middle range, all the distributions are asymmetric. This result is consistent with Figure 2(e) that $\kappa > 1$ when $I_G$ is relatively small, and $\kappa \to 1$ when $I_G$ is large.

A1.ii Assessment of the nominal crossing ratio

The setup is the same as Figure 10(a). The parameters and the governing equations of particle-gate collision are the same as in Section A1.i, except that $\psi$ varies from $\frac{\pi}{2} - \theta$ to $\frac{3\pi}{2} - \theta$.

For the forward process from left to right, in each calculation case, initially the particle is placed on the left-hand side of the dash-dotted circle, with given $v_i$, $\theta$, $\psi$, $\omega_i$, and $\phi$, where $v_i$ is the incident particle velocity and $\omega_i$ is the initial angular velocity of the gate. For a particle that crosses the line of opening of the diving wall (OC), if it collides with the gate, the collision location and the velocity of the outgoing particle are calculated. We count the total number of the cases that the particle can reach the right-hand side of the circle ($N_c$). For the reverse process from right to left, in each calculation case, initially the particle is placed on the right-hand side of the dash-dotted circle, with given $v_i$, $\theta$, $\psi$, $\omega_i$, and $\phi$. Based on the calculated particle trajectories, we count the total number of the cases that the line of opening (OC) is crossed ($N_c'$). The resolutions of $\theta$, $\psi$, $\phi$, $v_i$, and $\omega_i$ are 1/200, 1/400, 1/200, 1/15, and 1/15 of the full scales, respectively. The moment of inertia of the gate ($I_G$) ranges from $10^{-4}$ to $10^4$. For each $I_G$, more than $7.3 \times 10^9$ cases are simulated. The nominal crossing ratio is defined as $\kappa_o = \sum_1^{N_c} \hat{p}(v_i)\hat{p}(\omega_i) / \sum_1^{N'_c} \hat{p}(v_i)\hat{p}(\omega_i)$, where $\Sigma$ in the numerator and the denominator indicate summation for the forward crossing cases and the reverse crossing cases, respectively; $\hat{p}(\bullet)$ indicates the probability density; $\hat{p}(v_i)$ follows the two-dimensional Maxwell-Boltzmann distribution, and $\hat{p}(\omega_i)$ follows the one-dimensional Maxwell-Boltzmann distribution.

Figure 10(b) shows the computed $\kappa_o$ as a function of $I_G$. The trend qualitatively agrees with the MC simulation result (Figure 2e). For a light gate of a small $I_G$, $\kappa_o > 1$. For a heavy gate of a large $I_G$, $\kappa_o$ decreases to 1. The difference between Figure 10(b) and Figure 2(e) may be attributed to the large difference in particle size, as well as the possible multiple collisions in the gate zone that are ignored in this section.

**A2. Monte Carlo simulation**

The Monte Carlo (MC) simulation is carried out for a billiard-like particle, as shown in Figure 2(b). The system is two-dimensional. The upper container wall (AA′), the lower container wall (BB′), and the dividing wall (OD) are diffusive. From a diffusive wall, the magnitude of the reflected velocity follows the Maxwell-Boltzmann distribution; the direction of the reflected particle is random. The left and the right borders (AB and A′B′) are open, using periodic boundary condition. The gate is a specular line. It freely swings around point O, and the swinging motion is limited to the "+" side by the "door stopper" at point C. There is no long-range force among the particle, the walls, and the gate. The area enclosed by the middle line (CD) and the track of movement of the edge of the gate (the dash-dotted semicircular curve) is defined as the gate zone.

The setup is scalable; an example of the unit system can be based on g/mole, Å, fs, and K. The container is 60 in length and 20 in width; the particle mass is 1; the particle diameter is 1; the opening in the dividing wall (OC) is 10 in width; the time step is 0.05; temperature ($T$) is 400; the moment of inertia of the gate ($I_G$) is in the range from $10^{-1}$ to $10^3$; the gate length is 10; the average kinetic energy of the particle reflected from a diffusive wall is $k_B T$; the average kinetic energy of the gate reflected from the dividing wall or point C is $k_B T/2$.

Initially, the gate is closed, and the particle moves horizontally from left to right toward the middle point of the gate, with the initial velocity of $9.118\times10^{-4}$ per time step. The system is randomized for $10^8$ time steps. After the preparatory period, every time when the particle crosses the gate zone from left to right, $n_\pm$ increases by 1; every time when the particle crosses the gate zone from right to left, $n_\mp$ increases by 1.

Figure 2(c) shows the time profiles of $n_\pm/n_\mp$ ($I_G = 1$). The three dotted lines are the reference curves. For one reference curve, the gate is removed. For another reference curve, the "door stopper" (point C) is removed, so that the gate can swing freely at both sides of the middle line (CD). For the blue-colored reference curve, the gate is outward-swinging at the right-hand side of the middle line, while the particle number is increased to 100. In the no-gate reference case, a screen wall is attached to the upper container wall at A and A′ along the lateral boundary, to interrupt the trivial horizontal particle movement. The screen wall is a diffusive line, with the length of 5. In the multi-particle reference case, the particles are first evenly generated in the container. The initial particle velocity is randomly oriented, and the magnitude follows the Maxwell-Boltzmann distribution. The system is prepared for $10^5$ time steps, after which counting of $n_\pm$ and $n_\mp$ begins. For all the reference cases, the crossing event is defined as that a particle passes the line of opening in the dividing wall (OC). The simulation continues until the $n_\pm/n_\mp$ ratio stabilizes at the stead-state level.

Figure 2(d) shows the effect of the particle number ($\widehat{N}$). The setup is similar to the single-particle simulation (Figure 2b), except that multiple particles are placed in the container. At time zero, all the particles are evenly distributed; the velocity follows the Maxwell-Boltzmann distribution, and the direction is random. The maximum allowed swinging angle of the gate ($\phi$) is set to 120°, to minimize the trivial situation of unobstructed opening. After the initial randomization period, we count the total numbers of the crossing events at the middle line (OC) in both directions ($n_\pm$ and $n_\mp$), until the $n_\pm/n_\mp$ ratio reaches the steady state. For $\widehat{N} = $ 1, 10, 20, 30, 40, and 100, the initial randomization periods are 50, 7, 5.5, 1.5, 1, and 1 million time steps, respectively; the overall simulation times are 900, 40, 11, 10, 5, and 3 million time steps, respectively.

Figure 2(e) shows the single-particle steady-state $n_\pm/n_\mp$ ratio as a function of the moment of inertia of the gate ($I_G$), where $I_G = m_G L_G^2/3$ is varied by changing the gate mass ($m_G$); the gate size ($L_G$) is kept constant. The setup is the same as Figure 2(b). In Figure 2(d,e), for each data point, three simulations are performed; the error bars are calculated as the 90% confidence interval, $\pm 1.645\, \sigma_{SD}/\sqrt{3}$, where $\sigma_{SD}$ is the standard deviation.

Each single-particle simulation of the outward-swinging gate demonstrates a particle flux. It is worth noting that the particle flux is not caused by the occupied space of the gate. In the work on the classical Smoluchowski's trapdoor in [13], the excluded volume of the trapdoor leads to different nominal particle densities at the two sides, but the difference is only 1-2% and the crossing ratio is not affected; contrary to Figure 2(c-e), the trapdoor side has a lower particle count. In fact, for Figure 2(c,e), the gate zone is excluded from the data analysis of $n_\pm/n_\mp$, since the crossing events are counted at the boundary of the gate zone. Furthermore, Figure 2(e) suggests that the gate mass is an important parameter, opposite to the conclusion of the study in [13].

### A3. Experimental details

<u>A3.i Materials processing</u>

The experiment was performed on Toray UTC-82V polyamide (PA) microporous membranes obtained from Sterlitech. The membrane thickness was ~10 μm and the pore size was below 1 nm.[36,37] A membrane piece was sectioned by a razor blade, about 1.7 cm in diameter. It was attached to the stainless-steel inner frame of a McMaster-4518K63 compound o-ring (Figure 4d), using McMaster-7541A77 Devon epoxy. The epoxy was cured at room temperature for 24 h. The membrane was thoroughly cleaned by deionized (DI) water, and then immersed in 50 wt.% aqueous solution of isopropyl alcohol (IPA) for 24 h. Untreated membrane was dried at 75 °C for 30 min.

For the surface treatment, lauric acid (LA) and sulfuric acid ($H_2SO_4$) were provided by Sigma Aldrich (CAS No. 112-54-9 and CAS No. 7664-93-9, respectively). Similar to the procedure reported in [30], 20 mM aqueous solution of LA was prepared, and $H_2SO_4$ was dropped into it to adjust the pH value to 2. About 1 ml LA solution was added onto the front surface of the PA membrane, filling the steel frame (Figure 4e). The setup was heated at 75 °C for 30 min in a Jeio Tech ON-01E-120 oven. Then, the LA solution was removed and the membrane was repeatedly rinsed by DI water, immersed in DI water at 50 °C for 2 h, dried at 75 °C for 3 h, and rested at ambient temperature for 24 h. The Viton fluoroelastomer outer ring was placed onto the steel inner frame. The contact-angle measurement result confirmed that the grafting of dodecyl chains was successful (Figure 14).

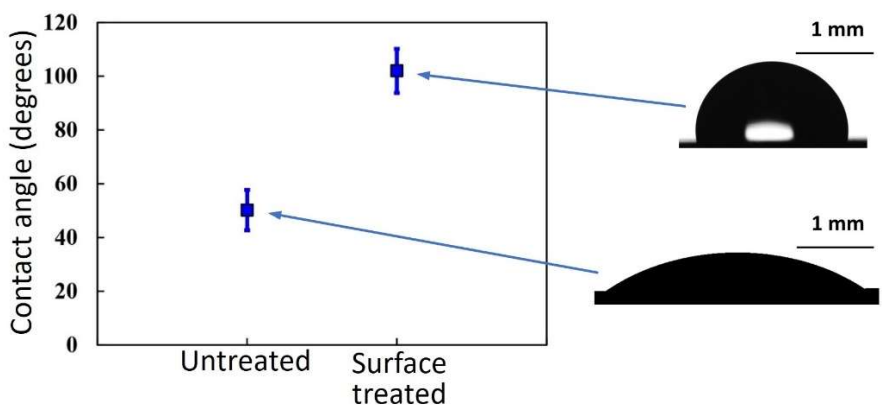

**Figure 14.** Contact angle measurement is a common technique to characterize surface-treated materials.[38,39] Upon surface treatment, the contact angle of a water drop on the polyamide membrane increased from ~50° to ~102°, indicating that the hydrophilic amide linkages were end-capped by the hydrophobic dodecyl chains. The insets on the right-hand side are photos of sessile water drops on a surface-grafted (top) and an untreated (bottom) polyamide membrane, observed through a ramé-hart model-200 contact angle goniometer at ambient temperature.

A3.ii Testing system

Figure 1(b), Figure 3, and Figure 15 show the testing system. The compound o-ring with a one-sidedly surface-grafted membrane was placed in between valves B1 and B2. An untreated

membrane was mounted on a similar compound o-ring, and placed in between valves C1 and C2. The gas phase was $C_2F_5I$. Table 1 lists the major system components. The containers mainly consisted of thin-walled stainless-steel vacuum hoses, four-way connectors, and flexible couplings, and were connected to a MTI EQ-FYP-Pump-110 vacuum pump, two Inficon SKY-CDG200D pressure sensors, and a $C_2F_5I$ gas storage vessel (Sigma Aldrich, CAS No. 354-64-3). The connections used vacuum clamps, o-rings, and vacuum grease.

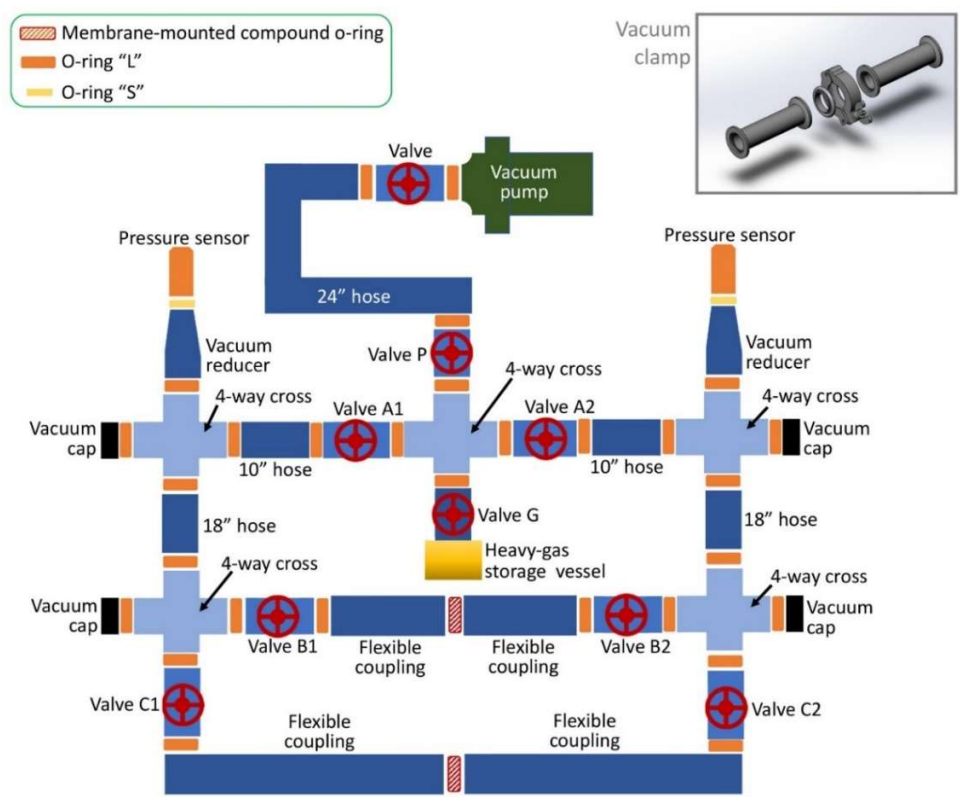

**Figure 15.** Schematic of the experimental setup. Vacuum clamps (see the inset at the upper-right corner) and vacuum grease are used at all the connections.

**Table 1** Parts list of the experimental system

| Vender | Part name | Description | Product number |
|---|---|---|---|
| **MTI** | Vacuum pump | UL certified 156 L/m double stage rotary vane vacuum pump with exhaust filter-EQ-FYP-Pump | EQ-FYP-Pump-110 |
| **Inficon** | Pressure sensor | SKY CDG200D capacitance diaphragm gauge | 3CF1-751-2300 |
| **Nor-Cal** | Vacuum valve | 1" manual angle valve, NW-25 flanges | ESV-1002-NWB |
| | Vacuum hose | NW-25 thin-wall stainless-steel flexible hose, with the length being 25.4 cm, 45.7 cm, or 61 cm. | LH-100-10-2NW<br>LH-100-18-2NW<br>LH-100-24-2NW |

| | Vacuum flexible coupling | NW-25 flexible coupling, 3.2" free length | 2FC-NW-25-1 |
|---|---|---|---|
| | Vacuum connector | NW-25 four-way cross connector | 4C-NW-25B |
| | Vacuum clamp | NW-25 wing nut clamp | NW-25-CP |
| **McMaster-Carr** | Vacuum o-ring "S" | O-Ring for 3/4" tube OD Quick-Clamp high-vacuum fitting | 4518K621 |
| | Vacuum o-ring "L" | O-Ring for 1" tube OD Quick-Clamp high-vacuum fitting | 4518K63 |
| | Vacuum cap | Cap for 1" stainless-steel tube OD Quick-Clamp high-vacuum fitting | 4518K58 |
| | Vacuum reducer | Quick-Clamp high-vacuum fitting | 4518K281 |
| **Dow Corning** | Vacuum grease | Dow Corning high-vacuum grease | 1597418 |

A3.iii System preparation and the initial condition

To prepare the system, valve G was closed, and all the other valves were open. The vacuum pump was turned on. The gas pressure was reduced to below 0.06 mTorr for 1 h. The pressure sensors were calibrated. Valve P was closed, and the pump was turned off. Valve G was opened, and $C_2F_5I$ slowly flew into the containers, until the pressure sensor readings reached ~0.8 Torr. The relatively low gas pressure had a low requirement on the membrane strength, and was within the pressure limit of the sensors. Valve G was closed, and the system rested for 2 h. If we needed to change the membrane, the valves across it would be closed and the low pressure was maintained in the rest of the system. After the membrane change, the operation of the vacuum pump was repeated.

For some tests (Figure 5f and Figure 8), the system temperature was first raised to 75 ºC. Valves G, C1, and C2 were shut, and all the other valves were open. The vacuum pump was turned on for 24 h, and the gas pressure was kept below 0.06 mTorr. Then, the system was cooled down to room temperature for 2 h. Valve P was closed, and valve G was opened, allowing $C_2F_5I$ to slowly flow into the containers, until $P_1$ and $P_2$ reached ~0.8 Torr. Valve G was closed, and the system rested for 30 min.

After the system preparation, valves A1, A2, B1, and B2 were open, and all the other valves remained shut. It was confirmed that at ~0.8 Torr, $C_2F_5I$ behaved as an ideal gas (Figure 16).

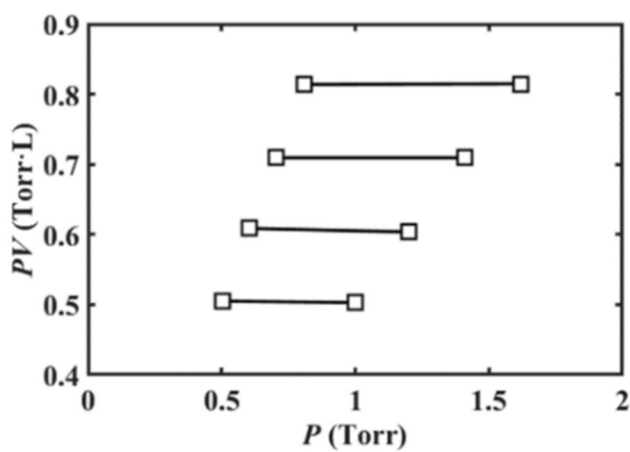

**Figure 16.** Based on Boyle's law, the pressure-volume measurement result supports the ideal-gas assumption. At ambient temperature, two identical chambers were separated by a closed vacuum valve. The chamber volume was 503 cm$^3$. The left chamber was filled by pentafluoroiodoethane ($C_2F_5I$) gas, and the initial gas pressure was around 1.0, 1.2, 1.4, or 1.6 Torr. The initial gas pressure in the right chamber was below 0.6 mTorr. Then, the valve was opened, and the pressure was measured again. Before and after the valve operation, the product of gas pressure ($P$) and gas volume ($V$) remained nearly constant, indicating that in such a low pressure range, $C_2F_5I$ can be analyzed as an ideal gas.[40,41]

Before the surface treatment, the gas permeability of each untreated membrane sample was observed. The procedure was similar to the measurement of Figure 5(b) (see Section A3.vi below), except that valves B1 and B2 remained closed and valves C1 and C2 remained open. The untreated membrane was installed between valves C1 and C2. The initial $\Delta P$ was set to ~2 mTorr. If the average decrease rate of $\Delta P$ was different from 5~10 µTorr/sec by more than 50%, the sample would be rejected. About 2/5 of the membrane sheets met this criterion.

A3.iv Measurement of the pressure difference: Figures 5(a,c)

Initially, $\Delta P \approx 0$. The surface-treated side of the membrane was toward container 2. We closed valves A1 and A2, leaving only valves B1 and B2 open. The operation of the two valves was steady and simultaneous, to minimize the disturbance on the pressure measurement. The readings of $P_1$ and $P_2$ were recorded. After ~10 min, valves A1 and A2 were opened again, followed by repeating the process for two more cycles, as shown by the black curve in Figure 5(a). Then, the membrane sample was flipped, so that the surface-treated side was toward container 1. The same $\Delta P$ measurement was performed, and the result is shown by the black curve in Figure 5(c).

The measurement procedures of the red and the gray curves in Figure 5(a) were similar, except that valves B1 and B2 remained shut and valves C1 and C2 remained open. For the gray curve, the untreated membrane in between valves C1 and C2 was replaced by a non-permeable 250 μm-thick solid polycarbonate film (McMaster 85585K103).

A3.v Effect of the initial pressure difference: Figures 5(b,d)

Initially, $P_1$ and $P_2$ were ~0.8 Torr. Valves B1 and B2 were open and all the other valves were closed. Then, valve P was opened, and the vacuum pump was turned on to reduce the pressure in the vacuum hose between valves A1 and A2. After ~5 sec., valve P was closed. Valve A1 or A2 were opened, so that the gas in container 1 or 2 flew into the section between valves A1 and A2, and $P_1$ or $P_2$ decreased by ~2 mTorr, causing an initial $\Delta P$. The changes of the pressure sensor readings were continuously monitored. Figures 5(b) and 5(d) show the measured $\Delta P$ profiles.

A3.vi Effect of the initial gas pressure: Figure 5(e)

Initially, the gas pressure ($P_1$ and $P_2$) was adjusted to about 0.4 Torr, 0.8 Torr, 1.2 Torr, or 1.6 Torr; valves A1 and A2 were closed. Valves B1 and B2 remained open; all the other valves remained shut. The testing procedure was similar to that of Figure 5(a). The result is given in Figure 5(e). For each initial pressure, three measurements were performed. The error bars were calculated as the standard deviation of the steady-state $\Delta P$.

A3.vii Variation of gas pressure in the two containers: Figure 5(f)

Initially, $P_1$ and $P_2$ were 801.5 mTorr; valves A1, A2, B1, B2 were open; all the other valves were closed. Then, valves A1 and A2 were shut. Valves B1 and B2 were slightly adjusted, to keep the gas pressure constant. The readings of sensor 1 ($P_1$) and sensor 2 ($P_2$) were continuously monitored. After $\Delta P$ has stabilized for ~15 min, valves A1 and A2 were opened again. The testing data indicate that, associated with the development of $\Delta P$ (Figure 5f, left), $P_1$ decreased by ~$\Delta P/2$ and $P_2$ increased by ~$\Delta P/2$ (Figure 5f, right).

A3.viii Negligible effect of gas adsorption/desorption of the membrane: Figure 7

Two 1.7 cm-diameter Toray UTC-82V polyamide (PA) microporous membranes and one 1.7 cm-diameter 250 μm-thick nonpermeable solid polycarbonate (PC) film (McMaster 85585K103) were harvested by a razor blade. One PA membrane was one-sidedly surface-treated. The untreated side of the PA membrane was firmly attached to the PC film by McMaster-7541A77 Devon epoxy. The other PA membrane was untreated. Its front surface was firmly attached to the other side of the PC film by the epoxy.

In Figure 1(b), the untreated membrane between valves C1 and C2 was replaced by the trilayer sample. Valves A1, A2, C1, and C2 were open; all the other valves remained closed. The initial gas pressure in containers 1 and 2 was ~0.8 Torr. The trilayer sample had the same front and back surfaces as the one-sidedly surface-grafted membrane, but the middle layer was nonporous, and no gas transport could take place.

At time zero, valves A1 and A2 were closed. Valves C1 and C2 were slightly adjusted to remove the disturbance on $\Delta P$. After about 25 min, valves A1 and A2 were reopened. Over time, little variation in pressure difference could be detected (Figure 7), suggesting that the effect of gas adsorption and desorption of the membrane was trivial.

A3.ix Effect of the gate-like chain behavior: Figure 8

Figure 8(a) shows the measured time profile of $\Delta P$ across an untreated membrane, with a dodecane layer physically adsorbed on one side. On a 1.3-cm$^2$ untreated Toray UTC-82V polyamide microporous membrane surface, ~1 ml 20-mM ethanol solution of dodecane (Sigma Aldrich, CAS No. 112-40-3) was dropped. The solvent was evaporated at 75 ºC in a Jeio Tech OV-12-120 oven at 20 kPa for 1 h, and then the membrane was rested at ambient temperature for 24 h. After the solvent deposition, the contact angle of water on the membrane increased from ~50º to ~100º, as shown by the inset in Figure 8(a). In Figure 1(b), the membrane in between valves B1 and B2 was replaced by the deposition-coated sample. The measurement procedure of $\Delta P$ was the same as that of Figure 5(a).

The upper curve in Figure 8(b) shows the pressure difference across an asymmetric bilayer sample. Two 1.7 cm-diameter Toray UTC-82V PA microporous membranes were harvested by a razor blade. One was one-sidedly surface-treated. The other membrane was not surface-treated. It was hydrothermally conditioned in a similar process, but no lauric aldehyde or acid was added in water. The two membranes were pressed together by a Vacmaster VP120 vacuum sealer. The surface-grafted side of the first membrane was attached to the front side of the second membrane. In Figure 1(b), the membrane in between valves B1 and B2 was replaced by the bilayer sample. The measurement procedure of $\Delta P$ was similar to that of Figure 5(a). It can be seen that addition of the untreated membrane has no significant influence on the steady-state $\Delta P$ of the one-sidedly surface-treated membrane.

The lower curve in Figure 8(b) shows the pressure difference across a symmetric bilayer sample. Two 1.7 cm-diameter Toray UTC-82V PA membranes were one-sidedly surface-treated. The untreated sides were firmly attached to each other by a Vacmaster VP120 vacuum sealer. In Figure 1(b), the membrane in between valves B1 and B2 was replaced by the bilayer sample. The measurement procedure was similar to that of Figure 5(a). The testing data showed that $\Delta P$ remained near zero.

Figure 8 suggests that $\Delta P$ is associated with the chemical bonding between the grafted dodecyl chains and the membrane surface. The covalent bonding enables the gate-like chain behavior. Physically adsorbed dodecane chains do not lead to a pressure difference.

A3.x Effect of the particle mass

For an order-of-magnitude assessment of the collision between a gas molecule and an organic chain, consider Figure 17. An elastic particle impacts a rigid rod. One end of the rod is hinged on the ground, and the other end is free. The rod can rotate around the hinge. The rod length is $L_0$, and its mass is $m_r$. Initially, the rod is vertical and its angular velocity is zero. The elastic particle moves in the horizontal direction, and impacts the rod at height $L_1$. The particle mass is $m_p$. After the rotational barrier of the hinge has been overcome, the effective incident velocity is $\hat{v}_0$. Upon collision, the particle velocity becomes $v_z$, and the rod gains an angular velocity $\omega$.

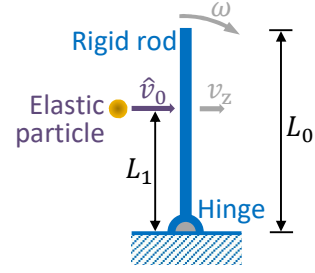

**Figure 17.** A particle collides with a vertical rod.

In accordance with the conservation of kinetic energy and angular momentum, we have $v_z = \frac{\varsigma_r - 1}{\varsigma_r + 1} \hat{v}_0$, where $\varsigma_r = \frac{3}{\hat{\varsigma}_r} \left(\frac{L_1}{L_0}\right)^2$ and $\hat{\varsigma}_r = \frac{m_r}{m_p}$. In order to keep $v_z$ and $\hat{v}_0$ in the same direction, $\varsigma_r$ needs to be larger than 1. Since $L_1 \leq L_0$, the upper bound of $\hat{\varsigma}_r$ is 3. That is, the particle mass must be at least 1/3 of the rod mass; otherwise, the particle would always be reflected.

**A4. Molecular dynamics simulation**

The molecular dynamics (MD) simulation was performed in LAMMPS.[42] The behavior of the dodecyl chain was simulated by the classical AIREBO potential.[43] The computer program is available at [14]. The system is three-dimensional. A (10,0) single-wall carbon nanotube (CNT) was employed as the analogue to a nanopore. The gas particle was a mercury (Hg) atom. The Hg atomic size is ~3 Å and its atomic mass ($m_{Hg}$) is 200.6.

The long-range interactions were described by the 12-6 Lennard-Jones potential, $\tilde{E} = 4\varepsilon_{\mu\nu} \left[\left(\frac{\sigma_{\mu\nu}}{r}\right)^{12} - \left(\frac{\sigma_{\mu\nu}}{r}\right)^{6}\right]$ (for $r \leq r_{co}$), where subscripts μ and ν represent atom type, $r$ is the atom-atom distance, $r_{co}$ is the cut-off distance, and $\varepsilon_{\mu\nu}$ and $\sigma_{\mu\nu}$ are two system parameters.

When $r > r_{co}$, $\tilde{E}$ was set to zero. The parameters of carbon-carbon, hydrogen-hydrogen, and mercury-mercury interactions have been well studied in open literature:[43,44] $\varepsilon_{CC} = 0.00284$ eV, $\sigma_{CC} = 3.4$ Å, $\varepsilon_{HH} = 0.0014994$ eV, $\sigma_{HH} = 2.65$ Å, $\varepsilon_{MM} = 0.0645$ eV, and $\sigma_{MM} = 2.969$ Å, where subscripts C, H, and M indicate carbon, hydrogen, and mercury, respectively. The Lorentz–Berthelot combining rule[45] was employed to compute the interactions between different types of atoms: $\sigma_{\mu\nu} = (\sigma_{\mu\mu} + \sigma_{\nu\nu})/2$ and $\varepsilon_{\mu\nu} = \sqrt{\varepsilon_{\mu\mu}\varepsilon_{\nu\nu}}$. For all the potentials, $r_{co}$ was 10.2 Å.

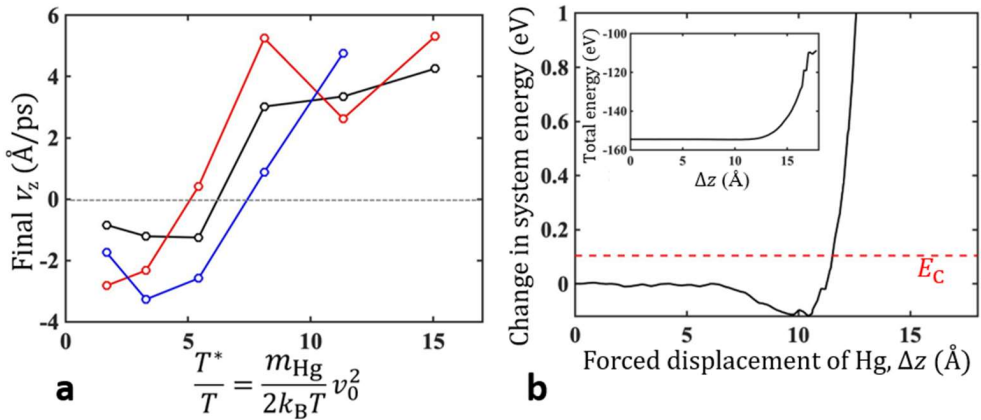

**Figure 18. (a)** The final z-dimension velocity of the Hg atom ($v_z$) as a function of the initial Hg velocity ($v_0$), for three randomly selected chain configurations. When $v_0$ is relatively low, the Hg atom cannot overcome the dodecyl chain and would be reflected back into the CNT, so that $v_z$ is negative. With a higher $v_0$, the Hg atom can push the chain open and exit the CNT, so that $v_z$ is positive. **(b)** A mercury (Hg) atom is forced to move downwards toward the CNT along $-z$, at a constant rate of 2 Å/ps. The variation in system energy is computed, as a function of the Hg displacement ($\Delta z$). The Hg atom starts at $\Delta z = 0$, 18 Å above the CNT opening. Initially, the system energy slightly decreases, because of the van der Waals attraction force between the chain and the Hg atom. When the chain is bent toward the CNT, as its motion is obstructed, the system energy drastically increases. The red dashed line indicates the level of the rotational barrier of carbon-carbon bond, $E_C \sim 0.1$ eV. The inset shows the overall profile of the system energy.

The length of the CNT was 17.04 Å and the inner diameter was 6.3 Å. At the end of the CNT, one 12-carbon hydrocarbon chain was covalently bonded. The unoccupied carbon atoms at the CNT edge were saturated by hydrogen. The CNT was placed at the center of a cuboid simulation box. In the x-y dimension normal to the CNT, periodic boundary condition was used. In the z direction along the CNT, the upper and bottom surfaces of the simulation box were isolated. The simulation box was 9.74 Å in width and 80 Å in length. The mercury (Hg) atom was placed on the center line of the CNT, either inside the CNT (17 Å below the CNT opening) or outside the CNT (18 Å above the CNT opening).

The time step was 0.1 fs. The carbon atoms in the CNT wall were fixed. The chain and the Hg atom were allowed to move. At 300 K, the system was equilibrated for 10 ps using the Langevin thermostat, and another 10 ps using the Nose-Hoover thermostat. Then, the temperature was raised to 1000 K in 20 ps with the Nose-Hoover thermostat. The high temperature was maintained for 10 ps, followed by cooling to 300 K in 10 ps and resting at 300 K for 10 ps. This process was repeated to randomly generate 100 chain configurations. They were used as the initial conditions for the study on the Hg-chain collision.

For each configuration, the Hg atom moved along the $z$ direction toward the dodecyl chain. The initial Hg velocity ($v_0$) ranged from 1.0 to 6.0 Å/ps. For Hg, $v_0 = 2$ Å/ps is the root mean square (rms) velocity at ~320 K. The simulation was stopped after 15 ps, or when the Hg atom crossed the lateral boundary of the simulation box. The final $z$-dimensional Hg velocity ($v_z$) and the final position of the Hg atom were recorded. The position was used to determine whether the Hg atom passed the chain. The probability of crossing was defined as $\delta_{cr} = N_{cr}/N_{tot}$, where $N_{tot} = 100$ is the total number of the investigated chain configurations and $N_{cr}$ is the number of passing cases (for a given $v_0$).

Figure 18(a) shows three examples of the final $z$-dimension Hg velocity ($v_z$). The Hg atom initially moves in the CNT upwards along $z$. When $T^* = \frac{m_{Hg}}{2k_B} v_0^2$ is low, Hg cannot overcome the rotational barrier of carbon-carbon bond ($E_C$) and is reflected back, so that $v_z$ is negative. When $T^*$ is large, $v_z$ becomes positive; i.e., the Hg atom crosses the CNT opening. The critical $T^*/T$ value of $v_z = 0$ is 5~7. The corresponding Boltzmann factor ($e^{-T^*/T}$) is 0.1%~0.6%. The first-order assessment of $\chi$ in Section 7 (~0.24%) is within this range. Figure 18(b) indicates that to force a Hg atom to push the chain into the CNT, the energy barrier is much higher than $E_C$, primarily because the CNT wall obstructs the chain movement.

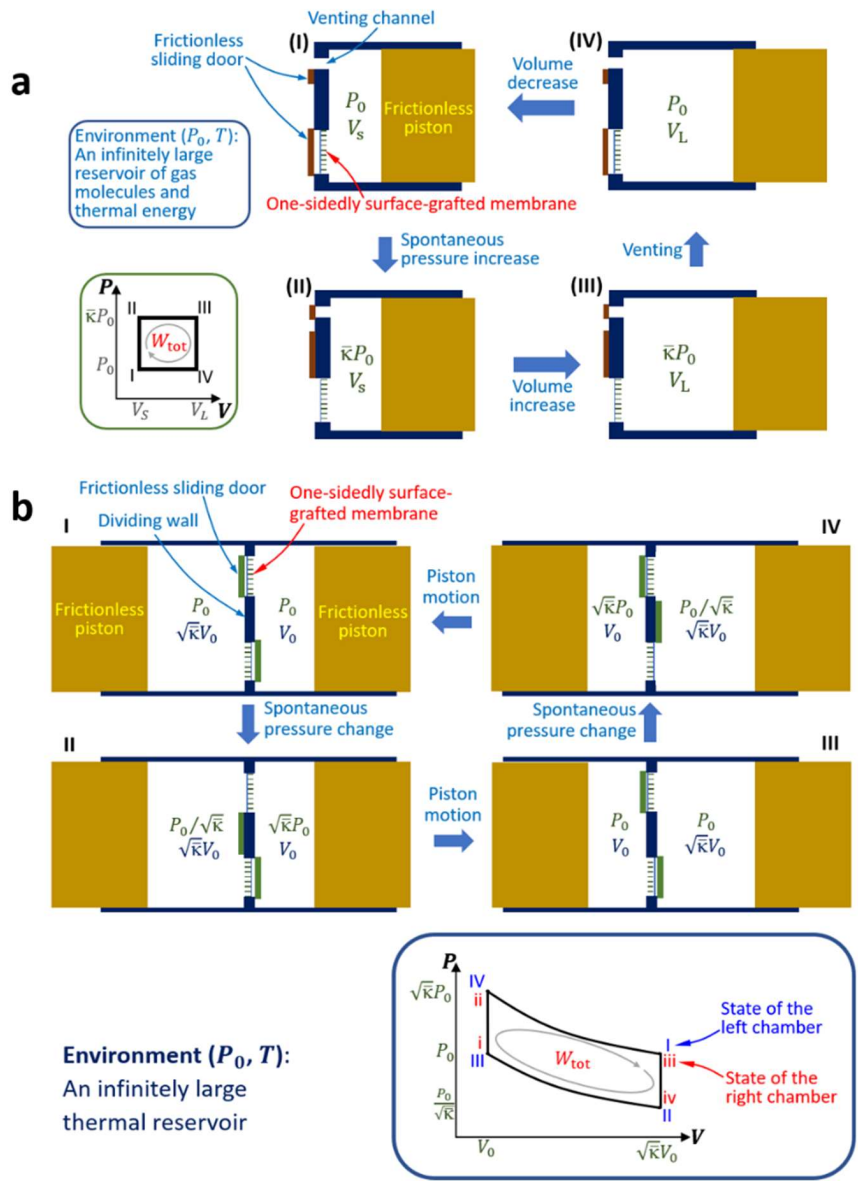

**Figure 19 (a)** Schematic of an isothermal cycle. In the pressure-volume $(P - V)$ loop at the lower-left corner, numbers I-IV indicate the system states. **(b)** A variant of the isothermal cycle, without mass exchange with the environment. In the pressure-volume $(P - V)$ loop at the lower-right corner, the blue upper-case numbers (I-IV) indicate the states of the left chamber; the red lower-case numbers (i-iv) indicate the states of the right chamber.

**A5. Production of useful work in an isothermal cycle**

In an isothermal setup, asymmetric membrane permeability ($\bar{\kappa} \neq 1$) cannot be described by the conventional theory of thermodynamics. It implies that useful work may be produced in a cycle by absorbing heat from a single thermal reservoir.

One example is given in Figure 19(a). The environment is a large reservoir of thermal energy and gas molecules, at constant pressure $P_0$ and temperature $T$. In the wall of a chamber, an asymmetric nanoporous membrane is installed. On the inner side of the membrane, there are molecular-sized outward-swinging gates at the nanopore openings. At State I, the membrane is covered by a frictionless sliding door, and the chamber is open to the environment through a regular venting channel, so that the inner gas pressure is $P_0$. The initial chamber volume is denoted by $V_S$. From State I to II, the venting channel is closed, and the membrane is exposed. The inner pressure spontaneously rises to $\bar{\kappa}P_0$. From State II to III, a frictionless piston moves out of the chamber and does work to the environment. The chamber volume increases to $V_L$, while the gas pressure remains $\bar{\kappa}P_0$. From State III to IV, the membrane is covered and the venting channel is open, and the inner pressure decreases to $P_0$. Finally, the piston moves back, and the chamber and the environment return to State I. After a complete cycle, the system produces work $W_{tot} = (\bar{\kappa} - 1)P_0 \Delta V$, where $\Delta V = V_L - V_S$. The produced work is from the absorbed heat from the environment.

Figure 19(b) shows another example, without mass transfer between the system and the environment. The environment is a large reservoir of thermal energy, at constant pressure ($P_0$) and temperature ($T$). There are two identical asymmetric nanoporous membranes in the dividing wall between two chambers filled with an ideal gas. Each membrane is one-sidedly grafted with molecular-sized outward-swinging gates. The gated side of the upper membrane faces the right chamber; the gated side of the lower membrane faces the left chamber. At State I, both membranes are covered by frictionless sliding doors. The gas pressure in the two chambers is the same $P_0$; the volumes of the left and the right chambers are $\sqrt{\bar{\kappa}}V_0$ and $V_0$, respectively. At State II, the upper membrane is exposed. As the gas diffuses from the left side to the right side across the upper membrane, the gas pressures in the left and the right chambers become $P_0/\sqrt{\bar{\kappa}}$ and $\sqrt{\bar{\kappa}}P_0$, respectively. Then, both membranes are covered again. The left piston moves into the left chamber, and the right piston moves out of the right chamber. At State III, the volumes of the left and the right chambers are $V_0$ and $\sqrt{\bar{\kappa}}V_0$, respectively. It can be seen that State III is symmetric to State I. The processes from State III to IV and from State IV to I are similar to I to II and II to III, respectively. For each cycle, the total input work is $W_{\text{in}} = P_0 V_0 \ln\bar{\kappa}$; the total output work is $W_{\text{out}} = P_0 V_0 \sqrt{\bar{\kappa}} \cdot \ln\bar{\kappa}$; the overall produced work is $W_{\text{tot}} = W_{\text{out}} - W_{\text{in}} = [(\sqrt{\bar{\kappa}} - 1) \cdot \ln\bar{\kappa}] P_0 V_0 = (\sqrt{\bar{\kappa}} - 1) W_{\text{in}}$.